\pdfoutput=1

\documentclass[12pt,a4paper]{article}

\usepackage{ifthen} 
\newboolean{pdflatex}
\setboolean{pdflatex}{true} 

\newboolean{articletitles}
\setboolean{articletitles}{true} 

\newboolean{uprightparticles}
\setboolean{uprightparticles}{false} 

\newboolean{inbibliography}
\setboolean{inbibliography}{false} 

\def\paperauthors{LHCb collaboration} 
\def\paperasciititle{ Measurement of the relative B- to D0/D*0/D**0 munu branching fractions using B- mesons from Bs2* decays} 
\def\papertitle{ Measurement of the relative \decay{\Bub}{\Dz / \Dstarz / \dstst \mun\neumb} branching fractions using \Bub mesons from \bstwostb decays} 
\def\paperkeywords{{High Energy Physics}, {LHCb}} 
\def\papercopyright{\the\year\ CERN on behalf of the LHCb collaboration}
\def\paperlicence{CC-BY-4.0 licence}
\def\paperlicenceurl{https://creativecommons.org/licenses/by/4.0/}

\usepackage[top=1in, bottom=1.25in, left=1in, right=1in]{geometry}

\usepackage{microtype}
\usepackage{lineno}  
\usepackage{xspace} 
\usepackage{caption} 

\usepackage{graphicx}  
\usepackage{color}
\usepackage{colortbl}
\graphicspath{{./figs_final/}} 

\usepackage{amsmath} 
\usepackage{amssymb}
\usepackage{amsfonts}
\usepackage{upgreek} 

\newcommand*\patchAmsMathEnvironmentForLineno[1]{%
\expandafter\let\csname old#1\expandafter\endcsname\csname #1\endcsname
\expandafter\let\csname oldend#1\expandafter\endcsname\csname
end#1\endcsname
 \renewenvironment{#1}%
   {\linenomath\csname old#1\endcsname}%
   {\csname oldend#1\endcsname\endlinenomath}%
}
\newcommand*\patchBothAmsMathEnvironmentsForLineno[1]{%
  \patchAmsMathEnvironmentForLineno{#1}%
  \patchAmsMathEnvironmentForLineno{#1*}%
}
\AtBeginDocument{%
\patchBothAmsMathEnvironmentsForLineno{equation}%
\patchBothAmsMathEnvironmentsForLineno{align}%
\patchBothAmsMathEnvironmentsForLineno{flalign}%
\patchBothAmsMathEnvironmentsForLineno{alignat}%
\patchBothAmsMathEnvironmentsForLineno{gather}%
\patchBothAmsMathEnvironmentsForLineno{multline}%
\patchBothAmsMathEnvironmentsForLineno{eqnarray}%
}

\usepackage{hyperxmp}

\usepackage[pdftex,
            pdfauthor={\paperauthors},
            pdftitle={\paperasciititle},
            pdfkeywords={\paperkeywords},
            pdfcopyright={Copyright (C) \papercopyright},
            pdflicenseurl={\paperlicenceurl}]{hyperref}

\usepackage[all]{hypcap} 

\usepackage{xspace} 
\usepackage{upgreek}

\def\lhcb   {\mbox{LHCb}\xspace}

\def\MagUp {\mbox{\em Mag\kern -0.05em Up}\xspace}

\ifthenelse{\boolean{uprightparticles}}%
{

 \def\Pmu         {\ensuremath{\upmu}\xspace}                 
 \def\Pnu         {\ensuremath{\upnu}\xspace}                 
                  
 \def\Ppi         {\ensuremath{\uppi}\xspace}

 \def\Ptau        {\ensuremath{\uptau}\xspace}

 \def\Ppsi        {\ensuremath{\uppsi}\xspace}

 \def\PDelta      {\ensuremath{\Delta}\xspace}                 
 \def\PXi      {\ensuremath{\Xi}\xspace}                 
 \def\PLambda      {\ensuremath{\Lambda}\xspace}                 
 \def\PSigma      {\ensuremath{\Sigma}\xspace}                 
 \def\POmega      {\ensuremath{\Omega}\xspace}                 
 \def\PUpsilon      {\ensuremath{\Upsilon}\xspace}                 
 
 \def\PB      {\ensuremath{\mathrm{B}}\xspace}                 
                  
 \def\PD      {\ensuremath{\mathrm{D}}\xspace}

 \def\PJ      {\ensuremath{\mathrm{J}}\xspace}                 
 \def\PK      {\ensuremath{\mathrm{K}}\xspace}

 \def\Pb      {\ensuremath{\mathrm{b}}\xspace}                 
 \def\Pc      {\ensuremath{\mathrm{c}}\xspace}

 \def\Pi      {\ensuremath{\mathrm{i}}\xspace}

 \def\Pp      {\ensuremath{\mathrm{p}}\xspace}

 \def\Ps      {\ensuremath{\mathrm{s}}\xspace}

}
{

 \def\Pmu         {\ensuremath{\mu}\xspace}                 
 \def\Pnu         {\ensuremath{\nu}\xspace}                 
                  
 \def\Ppi         {\ensuremath{\pi}\xspace}

 \def\Ptau        {\ensuremath{\tau}\xspace}

 \def\Ppsi        {\ensuremath{\psi}\xspace}                 
                  
 \mathchardef\PDelta="7101
 \mathchardef\PXi="7104
 \mathchardef\PLambda="7103
 \mathchardef\PSigma="7106
 \mathchardef\POmega="710A
 \mathchardef\PUpsilon="7107
                  
 \def\PB      {\ensuremath{B}\xspace}                 
                  
 \def\PD      {\ensuremath{D}\xspace}

 \def\PJ      {\ensuremath{J}\xspace}                 
 \def\PK      {\ensuremath{K}\xspace}

 \def\Pb      {\ensuremath{b}\xspace}                 
 \def\Pc      {\ensuremath{c}\xspace}

 \def\Pi      {\ensuremath{i}\xspace}

 \def\Pp      {\ensuremath{p}\xspace}

 \def\Ps      {\ensuremath{s}\xspace}

}

\makeatletter
\ifcase \@ptsize \relax
  \newcommand{\miniscule}{\@setfontsize\miniscule{4}{5}}
\or
  \newcommand{\miniscule}{\@setfontsize\miniscule{5}{6}}
\or
  \newcommand{\miniscule}{\@setfontsize\miniscule{5}{6}}
\fi
\makeatother

\DeclareRobustCommand{\optbar}[1]{\shortstack{{\miniscule (\rule[.5ex]{1.25em}{.18mm})}
  \\ [-.7ex] $#1$}}

\def\mup        {{\ensuremath{\Pmu^+}}\xspace}
\def\mun        {{\ensuremath{\Pmu^-}}\xspace} 

\def\taum       {{\ensuremath{\Ptau^-}}\xspace}

\def\neu        {{\ensuremath{\Pnu}}\xspace}
\def\neub       {{\ensuremath{\overline{\Pnu}}}\xspace}

\def\neumb      {{\ensuremath{\neub_\mu}}\xspace}
\def\neut       {{\ensuremath{\neu_\tau}}\xspace}
\def\neutb      {{\ensuremath{\neub_\tau}}\xspace}

\def\squark    {{\ensuremath{\Ps}}\xspace}

\def\cquark    {{\ensuremath{\Pc}}\xspace}

\def\bquark    {{\ensuremath{\Pb}}\xspace}

\def\pion   {{\ensuremath{\Ppi}}\xspace}

\def\pip    {{\ensuremath{\pion^+}}\xspace}
\def\pim    {{\ensuremath{\pion^-}}\xspace}

\def\kaon    {{\ensuremath{\PK}}\xspace}
  \def\Kbar    {{\kern 0.2em\overline{\kern -0.2em \PK}{}}\xspace}

\def\KorKbar    {\kern 0.18em\optbar{\kern -0.18em K}{}\xspace}

\def\Kp      {{\ensuremath{\kaon^+}}\xspace}
\def\Km      {{\ensuremath{\kaon^-}}\xspace}
\def\Kpm     {{\ensuremath{\kaon^\pm}}\xspace}

\def\Kstarb  {{\ensuremath{\Kbar{}^*}}\xspace}

  \def\Dbar    {{\kern 0.2em\overline{\kern -0.2em \PD}{}}\xspace}
\def\D       {{\ensuremath{\PD}}\xspace}

\def\DorDbar    {\kern 0.18em\optbar{\kern -0.18em D}{}\xspace}
\def\Dz      {{\ensuremath{\D^0}}\xspace}

\def\Dstar   {{\ensuremath{\D^*}}\xspace}

\def\Dstarz  {{\ensuremath{\D^{*0}}}\xspace}

\def\Dstarp  {{\ensuremath{\D^{*+}}}\xspace}

\def\B       {{\ensuremath{\PB}}\xspace}
\def\Bbar    {{\ensuremath{\kern 0.18em\overline{\kern -0.18em \PB}{}}}\xspace}
\def\Bb      {{\ensuremath{\Bbar}}\xspace}
\def\BorBbar    {\kern 0.18em\optbar{\kern -0.18em B}{}\xspace}

\def\Bub     {{\ensuremath{\B^-}}\xspace}

\def\Bs      {{\ensuremath{\B^0_\squark}}\xspace}
\def\Bsb     {{\ensuremath{\Bbar{}^0_\squark}}\xspace}

\def\Bdb     {{\ensuremath{\Bbar{}^0}}\xspace}

\def\jpsi     {{\ensuremath{{\PJ\mskip -3mu/\mskip -2mu\Ppsi\mskip 2mu}}}\xspace}

  \def\Y#1S{\ensuremath{\PUpsilon{(#1S)}}\xspace}

\def\proton      {{\ensuremath{\Pp}}\xspace}
\def\antiproton  {{\ensuremath{\overline \proton}}\xspace}

\def\Lz          {{\ensuremath{\PLambda}}\xspace}
\def\Lbar        {{\ensuremath{\kern 0.1em\overline{\kern -0.1em\PLambda}}}\xspace}
\def\LorLbar    {\kern 0.18em\optbar{\kern -0.18em \PLambda}{}\xspace}

\def\Lb      {{\ensuremath{\Lz^0_\bquark}}\xspace}

\def\BF         {{\ensuremath{\mathcal{B}}}\xspace}

\newcommand{\decay}[2]{\ensuremath{{#1\!\to #2}}\xspace}         

\def\to                 {\ensuremath{\rightarrow}\xspace}

\def\AT#1     {\ensuremath{A_{\mathrm{T}}^{#1}}\xspace}           

\def\C#1      {\ensuremath{\mathcal{C}_{#1}}\xspace}                       
\def\Cp#1     {\ensuremath{\mathcal{C}_{#1}^{'}}\xspace}                    
\def\Ceff#1   {\ensuremath{\mathcal{C}_{#1}^{\mathrm{(eff)}}}\xspace}        
\def\Cpeff#1  {\ensuremath{\mathcal{C}_{#1}^{'\mathrm{(eff)}}}\xspace}       
\def\Ope#1    {\ensuremath{\mathcal{O}_{#1}}\xspace}                       
\def\Opep#1   {\ensuremath{\mathcal{O}_{#1}^{'}}\xspace}                    

\newcommand{\tev}{\ifthenelse{\boolean{inbibliography}}{\ensuremath{~T\kern -0.05em eV}}{\ensuremath{\mathrm{\,Te\kern -0.1em V}}}\xspace}
\newcommand{\gev}{\ensuremath{\mathrm{\,Ge\kern -0.1em V}}\xspace}
\newcommand{\mev}{\ensuremath{\mathrm{\,Me\kern -0.1em V}}\xspace}
\newcommand{\kev}{\ensuremath{\mathrm{\,ke\kern -0.1em V}}\xspace}
\newcommand{\ev}{\ensuremath{\mathrm{\,e\kern -0.1em V}}\xspace}
\newcommand{\gevc}{\ensuremath{{\mathrm{\,Ge\kern -0.1em V\!/}c}}\xspace}
\newcommand{\mevc}{\ensuremath{{\mathrm{\,Me\kern -0.1em V\!/}c}}\xspace}
\newcommand{\gevcc}{\ensuremath{{\mathrm{\,Ge\kern -0.1em V\!/}c^2}}\xspace}
\newcommand{\gevgevcccc}{\ensuremath{{\mathrm{\,Ge\kern -0.1em V^2\!/}c^4}}\xspace}
\newcommand{\mevcc}{\ensuremath{{\mathrm{\,Me\kern -0.1em V\!/}c^2}}\xspace}

\def\mum  {\ensuremath{{\,\upmu\mathrm{m}}}\xspace}

\def\invfb   {\ensuremath{\mbox{\,fb}^{-1}}\xspace}

\def\gsim{{~\raise.15em\hbox{$>$}\kern-.85em
          \lower.35em\hbox{$\sim$}~}\xspace}
\def\lsim{{~\raise.15em\hbox{$<$}\kern-.85em
          \lower.35em\hbox{$\sim$}~}\xspace}

\def\pt         {\ensuremath{p_{\mathrm{ T}}}\xspace}
\def\ptot       {\ensuremath{p}\xspace}

\def\evtgen     {\mbox{\textsc{EvtGen}}\xspace}

\def\geant      {\mbox{\textsc{Geant4}}\xspace}

\def\photos     {\mbox{\textsc{Photos}}\xspace}

\def\pythia     {\mbox{\textsc{Pythia}}\xspace}

\def\tell1  {TELL1\xspace}
\def\ukl1   {UKL1\xspace}

\usepackage{subcaption}
\usepackage{siunitx}
\DeclareSIUnit\evolt{e\kern -0.1em V}
\usepackage{physics}
\usepackage{booktabs}
\usepackage[capitalize]{cleveref}
\crefname{section}{Sect.}{Sect.}
\Crefname{section}{Section}{Sections}

\usepackage{multirow}

\usepackage{cite} 
\usepackage{mciteplus}

\usepackage{morefloats}
\usepackage{longtable} 

\def\OSK {\text{OS}\kaon}
\def\SSK {\text{SS}\kaon}

\def\bsststb{{\ensuremath{\Bbar{}_s^{**0}}}\xspace}

\def\bstwostb{{\ensuremath{\Bbar{}_{s2}^{*0}}}\xspace}

\def\bsoneb{{\ensuremath{\Bbar{}_{s1}^0}}\xspace}

\def\dstst{{\ensuremath{D^{**0}}}\xspace}

\def\mmsq{\ensuremath{m_{\mathrm{miss}}^2}\xspace}
\def\mmin{\ensuremath{m_{\mathrm{min}}}\xspace}
\def\dmmin{\ensuremath{\Delta m_{\mathrm{min}}}\xspace}

\def\f12{\ensuremath{\frac{1}{2}}\xspace}

\begin{document}

\renewcommand{\thefootnote}{\fnsymbol{footnote}}
\setcounter{footnote}{1}



\begin{titlepage}
\pagenumbering{roman}

\vspace*{-1.5cm}
\centerline{\large EUROPEAN ORGANIZATION FOR NUCLEAR RESEARCH (CERN)}
\vspace*{1.5cm}
\noindent
\begin{tabular*}{\linewidth}{lc@{\extracolsep{\fill}}r@{\extracolsep{0pt}}}
\ifthenelse{\boolean{pdflatex}}
{\vspace*{-1.5cm}\mbox{\!\!\!\includegraphics[width=.14\textwidth]{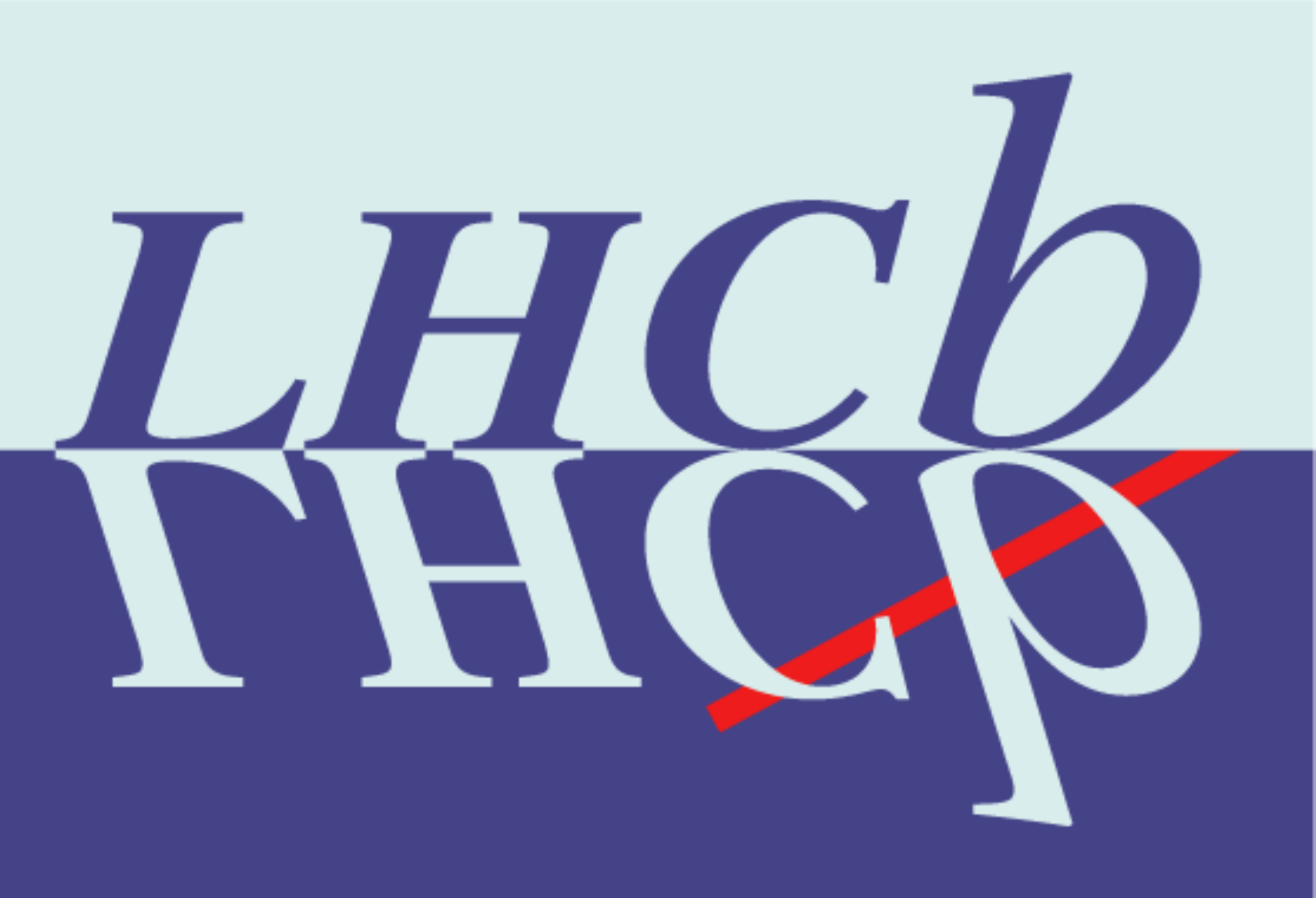}} & &}%
{\vspace*{-1.2cm}\mbox{\!\!\!\includegraphics[width=.12\textwidth]{lhcb-logo.eps}} & &}%
\\
 & & CERN-EP-2018-190 \\  
 & & LHCb-PAPER-2018-024 \\  
 & & 3 June, 2019 \\ 
\end{tabular*}

\vspace*{2.0cm}

{\normalfont\bfseries\boldmath\huge
\begin{center}
  \papertitle 
\end{center}
}

\vspace*{1.0cm}

\begin{center}
\paperauthors\footnote{Authors are listed at the end of this paper.}
\end{center}

\vspace{\fill}

\begin{abstract}
  \noindent
The decay of the narrow resonance \decay{\bstwostb}{\Bub\Kp} can be used to determine the \Bub momentum in partially reconstructed decays without any assumptions on the decay products of the \Bub meson.
   This technique is employed for the first time to distinguish contributions from \Dz, \Dstarz, and higher-mass charmed states (\dstst) in semileptonic \Bub decays by using the missing-mass distribution. The measurement is performed using a data sample corresponding to an integrated luminosity of $3.0\invfb$ collected with the LHCb detector in $pp$ collisions at center-of-mass energies of 7 and 8\tev.  The resulting branching fractions relative to the inclusive \decay{\Bub}{\Dz X \mun \neumb} are
  \begin{align*}
    f_{\Dz} = \BF\qty( \Bub \to \Dz\mun\neumb )/\BF\qty( \Bub \to \Dz X \mun\neumb ) &= 0.25 \pm 0.06, \\
    f_{\dstst}  = \BF\qty( \Bub \to \qty(\dstst\to\Dz X)\mun\neumb )/\BF\qty( \Bub \to \Dz X \mun\neumb ) &= 0.21 \pm 0.07,
  \end{align*}
  with $f_\Dstarz = 1 - f_\Dz - f_\dstst$ making up the remainder.
\end{abstract}


\vspace*{1.0cm}

\begin{center}
  Published in Phys.~Rev.~D99 (2019) 092009
\end{center}

\vspace{\fill}

{\footnotesize 
\centerline{\copyright~\papercopyright. \href{\paperlicenceurl}{\paperlicence}.}}
\vspace*{2mm}

\end{titlepage}


\newpage
\setcounter{page}{2}
\mbox{~}
%
%
%
%

\cleardoublepage


\renewcommand{\thefootnote}{\arabic{footnote}}
\setcounter{footnote}{0}



\pagestyle{plain} 
\setcounter{page}{1}
\pagenumbering{arabic}



\section{Introduction}
\label{sec:intro}

The composition of the inclusive bottom-to-charm semileptonic rate is not fully understood.  Measurements of the exclusive branching fractions for \decay{\B}{D\ell\nu} and \decay{\B}{\Dstar\ell\nu} and corresponding decays with up to two additional charged pions~\cite{HFLAV16} do not saturate the total \decay{b}{c} semileptonic rate as determined from analysis of the charged lepton's kinematic moments~\cite{Mahmood:2004kq,Aubert:2006au,Urquijo:2006wd}.
One way to resolve this inclusive--exclusive gap is to make measurements of relative rates between different final states. 

Semileptonic decays with excited charm states act as important backgrounds both to the exclusive decay channels \decay{B}{D\ell\nu} and \decay{B}{\Dstar\ell\nu} and for the study of semileptonic \decay{b}{u} transitions. 
For example, understanding these backgrounds is essential for experimental tests of lepton flavor universality studied by comparing the rates of tauonic and muonic $b$-hadron decays, \emph{e.g.} $R(D^{(*)}) \equiv \BF\qty(\decay{\Bbar}{D^{(*)}\taum\neutb})/\BF\qty(\decay{\Bb}{D^{(*)}\mun\neumb})$~\cite{Lees:2013uzd,Huschle:2015rga,LHCb-PAPER-2015-025,Sato:2016svk,Hirose:2016wfn,LHCb-PAPER-2017-017,LHCb-PAPER-2017-027}.\footnote{The inclusion of charge-conjugate processes is implied throughout.}

The largest contributions of excited charm states besides the $D^{*}(2007)^0$ or $D^{*}(2010)^+$ mesons come from the orbitally excited $L=1$ states $D_0^*\qty(2400)$, $D_1\qty(2420)$, $D_1\qty(2430)$, and $D_2^*\qty(2460)$, which have been individually measured~\cite{HFLAV16}. We use the collective term $D^{**}$ to refer to these as well as other resonances such as radially excited $D$ mesons, and to nonresonant contributions with additional pions.

The contribution of excited states to the total semileptonic rate can be studied using $B$ decays in which the $B$ momentum is known. This allows one to calculate the mass of the undetected or ``missing'' part of the decay, and thus separate different excited $D$ states.
In this paper we employ for the first time the technique described in Ref.~\cite{Stone:2014mza} to accomplish this reconstruction in \decay{\Bub}{\Dz X\mun\neumb} decays, where $X$ refers to any number of additional particles, without assumptions about the decay products of the \Bub meson. There are three narrow peaks in the $\Bub\Kp$ mass distribution just above the mass threshold from decays of the orbitally excited $L=1$ $\Bbar_s^{**}$ mesons~\cite{Aaltonen:2007ah,Abazov:2007af,LHCb-PAPER-2012-030}. We focus on the decay \decay{\bstwostb}{\Bub\Kp}, which forms a narrow peak approximately \SI{67}{\mega\evolt} above the threshold,\footnote{Natural units with $c = 1$ are used throughout.} and has the largest yield of any observed excited \Bsb state. By tagging \Bub mesons produced from the decay of these excited \bstwostb mesons, the \Bub energy can be determined up to a quadratic ambiguity using the \bstwostb and \Bub decay vertices and by imposing mass constraints for the \Bub and \bstwostb mesons. Since only approximately 1\% of \Bub mesons originate from a \bstwostb decay, this method requires a large data set.

We determine the relative branching fractions of \Bub to \Dz, \Dstarz, and \dstst, referred to as $f_\Dz$, $f_\Dstarz$, and $f_\dstst$ respectively, in the $\decay{\Bub}{\Dz X\mun \neumb}$ channel by fitting the distribution of the missing mass for \decay{\bstwostb}{\Bub\Kp} candidates.
A similar set of fractions (along with their \Bdb counterparts), where the charge of the final state $D$ meson is not specified, has been measured previously at the BaBar experiment~\cite{Aubert:2007bq}.
From the derivations in Ref. \cite{mynote}, we expect based on previous branching fraction measurements
\begin{align*}
f_{\Dz} &= \BF\qty( \Bub \to \Dz\mun\neumb )/\BF\qty( \Bub \to \Dz X \mun\neumb )     &  &= 0.235 \pm 0.011 ^{+0.018}_{ -0.012 }, \\
f_{\Dstarz} &= \BF\qty( \Bub \to \Dstarz\mun\neumb )/\BF\qty( \Bub \to \Dz X \mun\neumb ) &    &= 0.564 \pm 0.017 ^{+0.042}_{ -0.028 }, \\ 
f_{\dstst}  &= \BF\qty( \Bub \to \qty(\dstst\to\Dz X)\mun\neumb )/\BF\qty( \Bub \to \Dz X \mun\neumb ) &   &= 0.201 \pm 0.020 ^{+0.039}_{ -0.060 },  
\end{align*}
where the first uncertainty is experimental and the second gives an envelope of different extrapolation hypotheses to explain the inclusive--exclusive gap. Precise measurements of the relative branching fractions can distinguish between the hypotheses.
Higher values in the \dstst envelope (20\% or more) would point towards a scenario in which there is a large contribution of unmeasured excited charm states. Lower fractions, closer to 14\%, would suggest that the currently measured exclusive decays correctly describe the makeup of the total rate, and the inclusive--exclusive gap is due to other systematic effects.

A description of the data samples and selections used in this paper may be found in \cref{sec:select}. Afterwards we discuss the missing mass reconstruction and related variables in \cref{sec:reco}.
Along with the signal \bstwostb decays, a large fraction of background decays are also selected. Yields and missing mass shapes must be determined for each of the background categories as described in \cref{sec:backgrounds}. The most important background source is semileptonic decays of \Bub and \Bdb mesons with the same final state as the signal that do not originate from \bstwostb decays. After accounting for other sources of background in \cref{subsec:smallbkg}, we estimate the yield and shape of this source in \cref{subsec:slbkg}.
The relative branching fractions are determined using a template fit to the missing mass distribution as described in \cref{sec:fit}. The systematic uncertainties included in the fit are then described in \cref{sec:syst}. The final result is presented in \cref{sec:results}.

\section{Data sample and selection}
\label{sec:select}

The \lhcb detector~\cite{Alves:2008zz,LHCb-DP-2014-002} is a single-arm forward
spectrometer covering the \mbox{pseudorapidity} range $2<\eta <5$,
designed for the study of particles containing \bquark or \cquark
quarks. The detector includes a high-precision tracking system
consisting of a silicon-strip vertex detector surrounding the $pp$
interaction region~\cite{LHCb-DP-2014-001}, a large-area silicon-strip detector located
upstream of a dipole magnet with a bending power of about
$4{\mathrm{\,Tm}}$, and three stations of silicon-strip detectors and straw
drift tubes~\cite{LHCb-DP-2013-003} placed downstream of the magnet.
The tracking system provides a measurement of momentum, \ptot, of charged particles with
a relative uncertainty that varies from 0.5\% at low momentum to 1.0\% at 200\gev.
The minimum distance of a track to a primary vertex (PV), the impact parameter (IP), 
is measured with a resolution of $(15+29/\pt)\mum$,
where \pt is the component of the momentum transverse to the beam, in\,\gev.
Different types of charged hadrons are distinguished using information
from two ring-imaging Cherenkov detectors~\cite{LHCb-DP-2012-003}. 
Photons, electrons and hadrons are identified by a calorimeter system consisting of
scintillating-pad and preshower detectors, an electromagnetic
calorimeter and a hadronic calorimeter. Muons are identified by a
system composed of alternating layers of iron and multiwire
proportional chambers~\cite{LHCb-DP-2012-002}.
The online event selection is performed by a trigger~\cite{LHCb-DP-2012-004}, 
which consists of a hardware stage, based on information from the calorimeter and muon
systems, followed by a software stage, which applies a full event
reconstruction.

We use data samples collected in 2011 and 2012, at center-of-mass energies of \SI{7}{\tera\evolt} and \SI{8}{\tera\evolt} respectively, corresponding to an integrated luminosity of 3.0\invfb.  All \Bub candidates are selected from $\Dz\mun$ combinations, with \decay{\Dz}{\Km\pip}.  The final-state particles are formed from high-quality tracks required to be inconsistent with being produced at any primary collision vertex in the event. Loose particle-identification requirements are also applied to these tracks. The $\Km$ and $\pip$ candidates must form a high-quality vertex, and their combined mass must lie in the range \SIrange{1840}{1890}{\mega\evolt}. The muon from the $\Dz\mun$ candidate is required to pass the hardware trigger, which requires a transverse momentum of $\pt > \SI{1.48}{\giga\evolt}$ in the \SI{7}{\tera\evolt} data or $\pt > \SI{1.76}{\giga\evolt}$ in the \SI{8}{\tera\evolt} data. The software trigger requires a two-, three- or four-track secondary vertex with a significant displacement from
any primary $pp$ interaction vertex, consistent with coming from a $b$ hadron.  The $\Dz\mun$ vertex must be of high quality, and well separated from the primary vertex.

After selecting \Bub candidates, we add candidate kaons consistent with originating from the primary vertex, referred to as prompt, to form the \bstwostb candidates.  To reduce background from misidentified pions from the primary interaction, we impose strong particle-identification requirements.  The selection requirements for the prompt kaons are optimized using the fully reconstructed decay \decay{\Bub}{\jpsi \Km}. Signal decays produce a $\Bub\Kp$ pair; in addition to this opposite-sign kaon (\OSK) data sample, we also use $\Bub\Km$ same-sign kaon (\SSK) combinations to help estimate backgrounds from data.

Samples of simulated \bstwostb events are used to model the  \decay{\Bub}{\Dz\mun\neumb}, \decay{\Bub}{\Dstarz\mun\neumb}, and \decay{\Bub}{\dstst\mun\neumb} signal components. For the \dstst component, the simulation includes contributions from the four $L=1$ $D$ mesons as well as a small contributions of nonresonant $D^{(*)}\pi$ decays.
In the simulation, $pp$ collisions are generated using
\pythia~\cite{Sjostrand:2006za,*Sjostrand:2007gs} 
 with a specific \lhcb
configuration~\cite{LHCb-PROC-2010-056}.  Decays of hadronic particles
are described by \evtgen~\cite{Lange:2001uf}, in which final-state
radiation is generated using \photos~\cite{Golonka:2005pn}. The
interaction of the generated particles with the detector, and its response,
are implemented using the \geant
toolkit~\cite{Allison:2006ve, *Agostinelli:2002hh} as described in
Ref.~\cite{LHCb-PROC-2011-006}.

\section{Reconstruction of the $\boldsymbol{\Bub}$ meson momentum}
\label{sec:reco}

\begin{figure}[tb]
  \includegraphics[width=\textwidth]{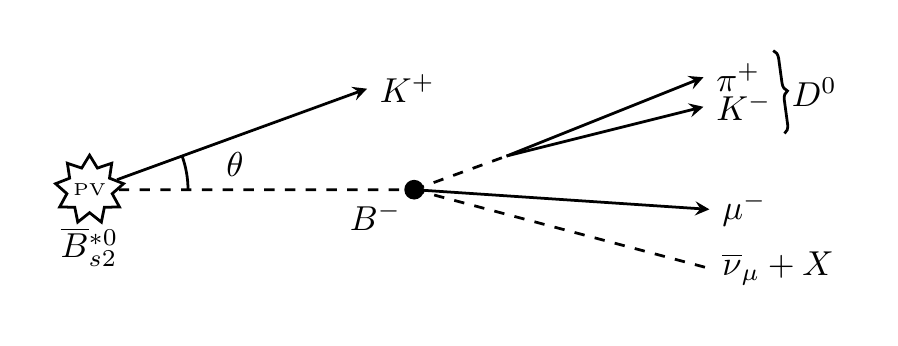}
  \caption{Decay topology for the \decay{\Bub}{\Dz X \mun \neumb} signal decays. A \bstwostb meson decays at the primary vertex position, producing a \Bub meson and a \Kp meson. The angle in the laboratory frame between the \Kp and \Bub directions is defined as $\theta$. The \Bub meson then decays semileptonically to a \Dz meson and a muon, accompanied by an undetected neutrino and potentially other particles, referred to collectively as $X$.\label{fig:decay_chain}}
\end{figure}

We find the energy of the \Bub meson by using its flight direction from the primary vertex to the secondary $\Dz\mun$ vertex; a diagram of the decay topology is shown in \cref{fig:decay_chain}.  Applying mass constraints for the \Bub meson mass, $m_B$, and the hypothesized parent particle mass, $m_{BK}$, leaves a quadratic equation for the \Bub meson energy, $E_B$, derived in \cref{app:eq}. 

\begin{figure}[tbp]
  \begin{center}
    \begin{subfigure}{0.7\textwidth}
      \includegraphics[width=\textwidth]{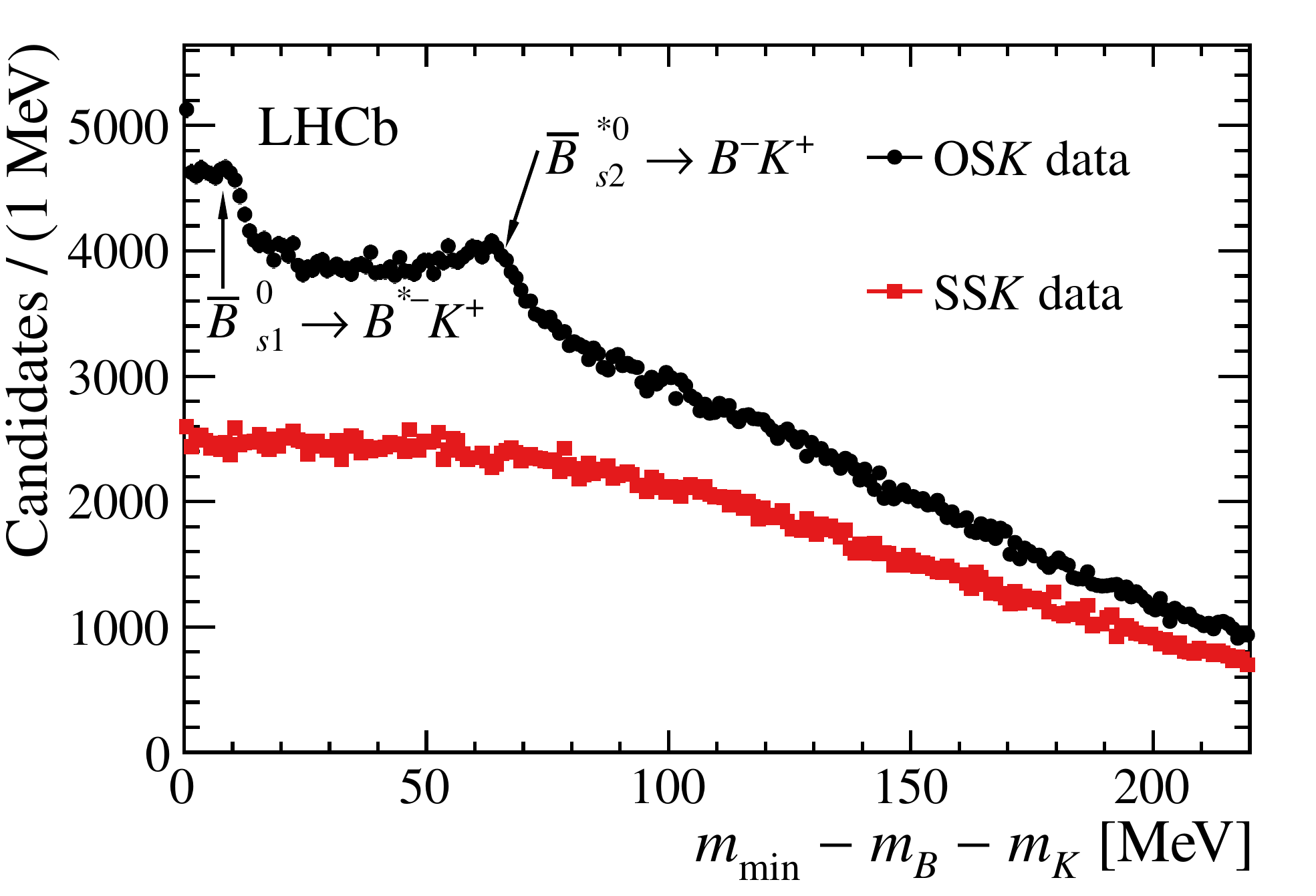}
    \end{subfigure}
  \end{center}
  \caption{
    Distribution of the minimum mass difference for $\Bub\Kp$ (\OSK) candidates and $\Bub\Km$ (\SSK) candidates. For \OSK combinations, peaks for the \bstwostb and \bsoneb states are visible. The contribution of decays in which a kaon from a $b$-hadron decay is chosen as prompt produces the sharp increase near zero.  The \SSK sample is used for background estimation.\label{fig:mmin}
  }
\end{figure}

 In carrying out the analysis we use two different quantities related to this calculation.
The first is the minimum mass of the $\Bub\Kpm$ pair. For a particular \Bub vertex and kaon track, there is a minimum $m_{BK}$ mass hypothesis for which the \Bub energy solutions are real. At this value, the discriminant of the quadratic equation is zero. This minimum mass value is given by
\begin{equation}
  \mmin = \sqrt{ m_B^2 + m_K^2 + 2m_B\sqrt{ p_K^2\sin^2\theta + m_K^2} },
\end{equation} 
where $p_K$ is the kaon momentum in the laboratory frame, $m_K$ is the kaon mass, and $\theta$ is the angle between the kaon direction and the direction from the primary to the secondary vertex.  The distribution of the difference between \mmin and the $m_B + m_K$ threshold, $\dmmin = \mmin - m_B - m_K$,  shown in \cref{fig:mmin} for both the \OSK and \SSK data samples, has excesses corresponding to the \bstwostb and \bsoneb states even for decays that are not fully reconstructed.  We use these distributions in a control region of $ 0 < \dmmin < \SI{220}{\mega\evolt}$ to constrain the total amount of \bstwostb decays and non-\bstwostb background contributions in our selection, as described in more detail in \cref{sec:backgrounds}.

Decays of \bsoneb mesons and background candidates where a secondary kaon is misidentified as coming from the primary interaction have small values of \dmmin; the latter produces the increase near zero seen in \cref{fig:mmin}. To remove these, we define our signal region for the missing mass fit as $ 30 < \dmmin < \SI{67}{\mega\evolt}$.

The second quantity is the missing mass, assuming the particles result from the decay of a \bstwostb meson (imposing $m_{BK} = m_{\bstwostb}$). The energy of the \Bub meson, $E_B$, is calculated as follows:
\begin{align}
  E_B &= \frac{\Delta^2}{2E_K}\frac{1}{1 - \qty(p_K/E_K)^2 \cos^2\theta} \qty[ 1 \pm \sqrt{d} ], \label{eqn:eb} \\
  \intertext{where}
  \Delta^2 &= m_{\bstwostb}^2 - m_B^2 - m_K^2, \\
  \intertext{and}
  d &= \frac{p_K^2}{E_K^2} \cos^2\theta - \frac{4m_B^2p_K^2\cos^2\theta}{\Delta^4}\qty(1 - \frac{p_K^2}{E_K^2} \cos^2\theta ).
\end{align}
Once $E_B$ has been determined, we calculate the missing mass squared
\begin{equation}
  \mmsq = ( p_B - p_{\mathrm{vis}} )^2,
\end{equation}
where $p_B$ is the four momentum calculated from $E_B$ and the \Bub direction, and $p_{\mathrm{vis}}$ is the four momentum of the $\Dz\mun$ combination. We require real solutions for \cref{eqn:eb}. This keeps only candidates with \mmin less than the \bstwostb mass; candidates with $\dmmin > m_{\bstwostb}  - m_B - m_K$, which is approximately \SI{67}{\mega\evolt}, produce imaginary solutions. The \mmsq variable is then used to perform the final fit to determine the relative branching fractions.

\begin{figure}[tbp]
    \begin{subfigure}{0.5\textwidth}
      \includegraphics[width=\textwidth]{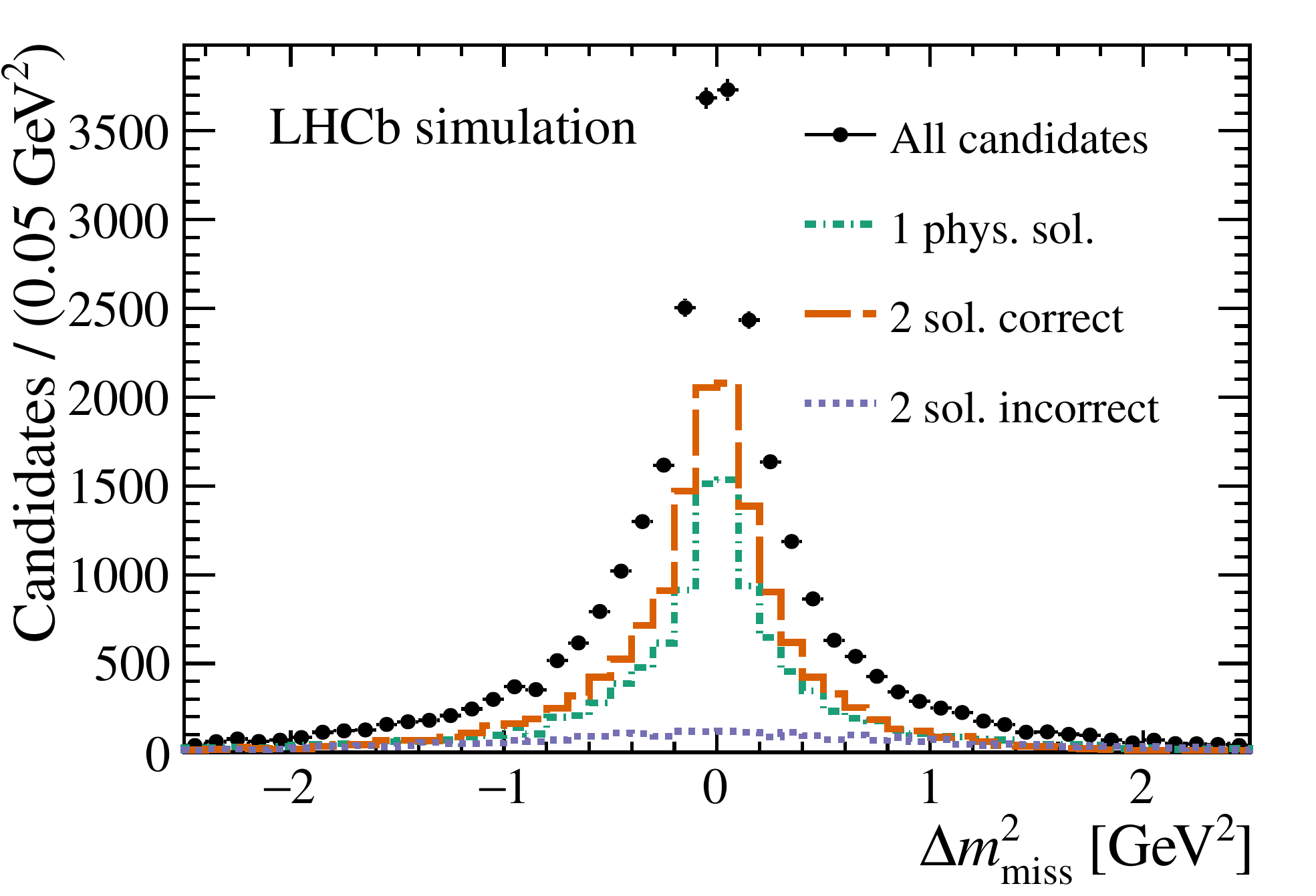}
    \end{subfigure}
    \begin{subfigure}{0.5\textwidth}
      \includegraphics[width=\textwidth]{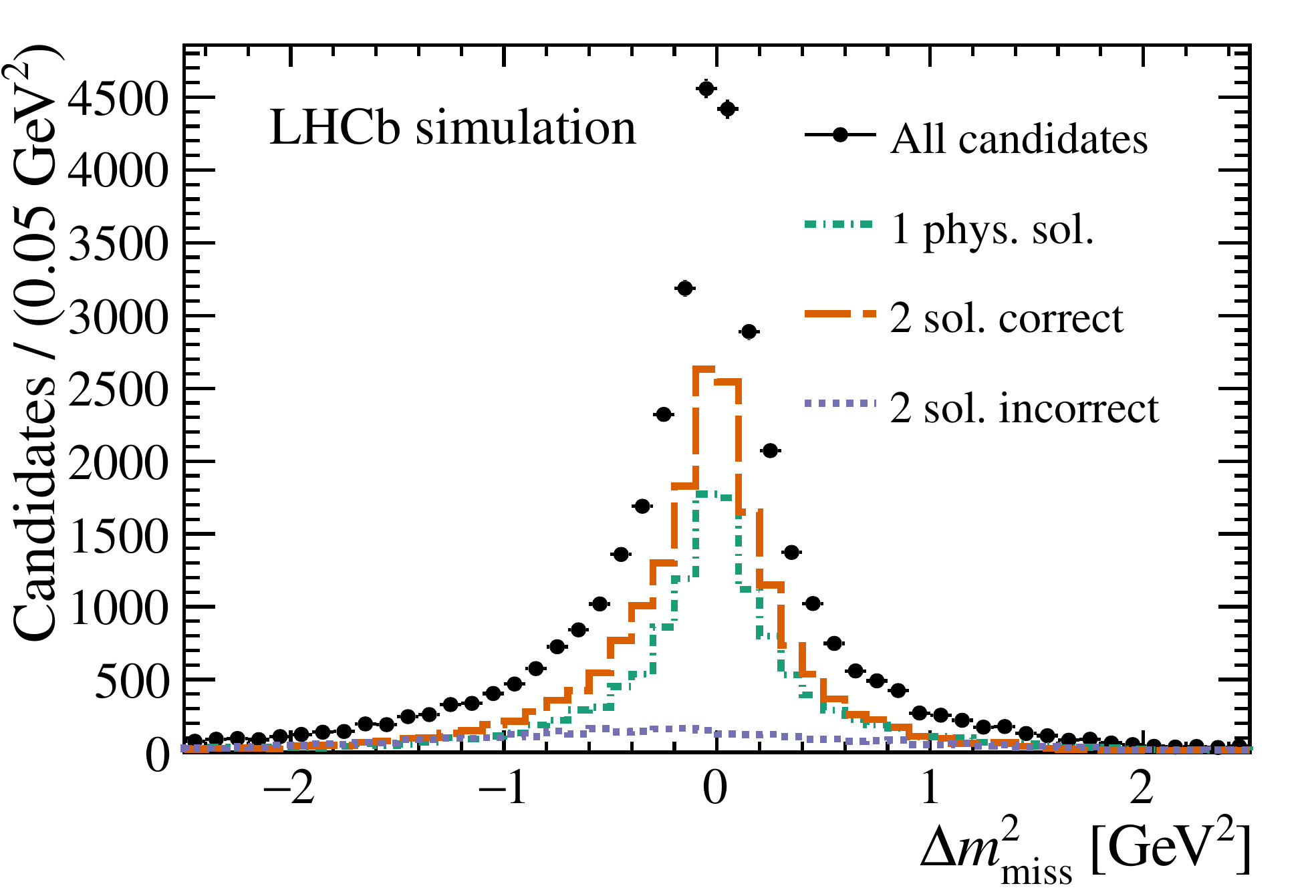}
    \end{subfigure} \\
  \begin{center}
    \begin{subfigure}{0.5\textwidth}
      \includegraphics[width=\textwidth]{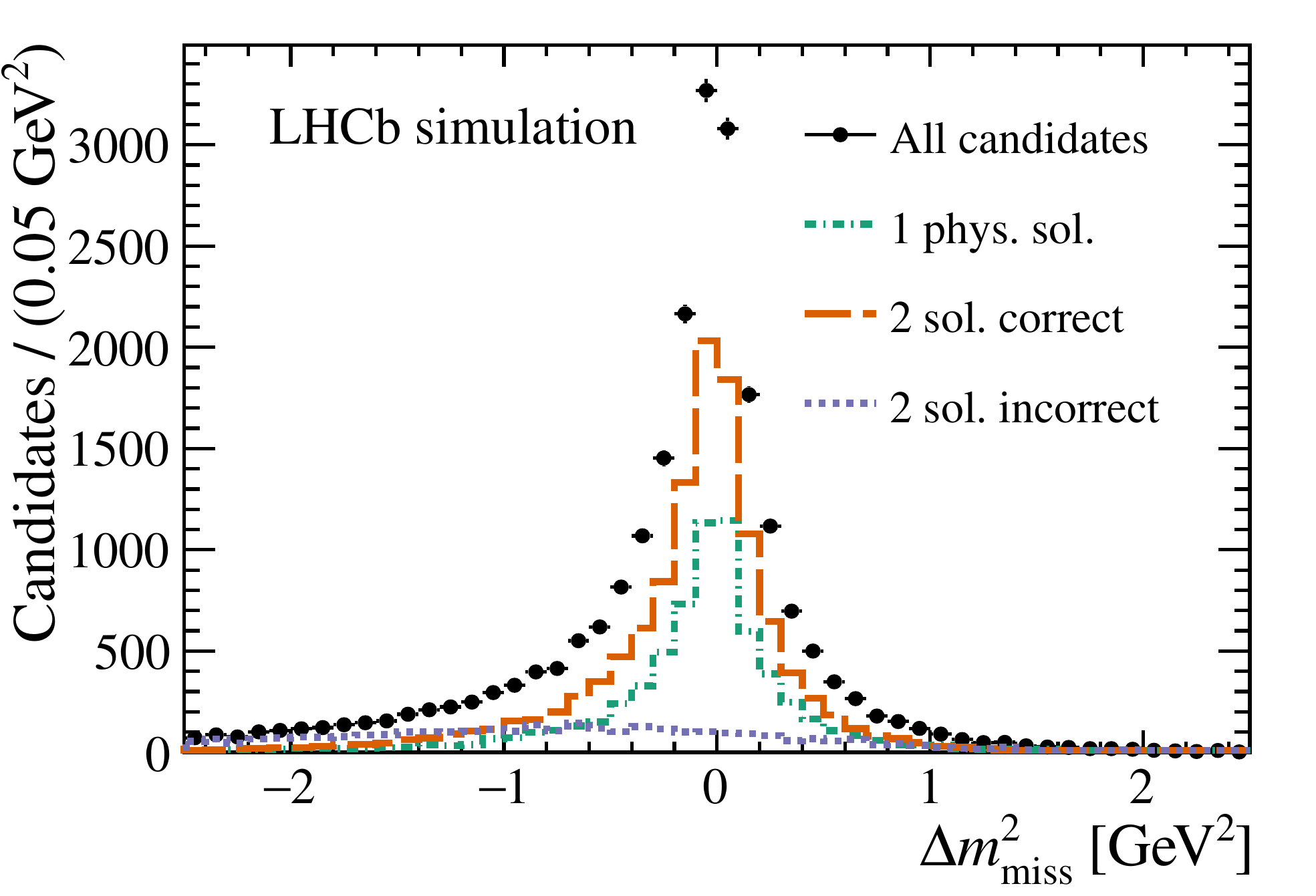}
    \end{subfigure}
  \end{center}
  \caption{The difference $\Delta\mmsq$ between the reconstructed missing mass squared and the corresponding true values for the (top left) \decay{\Bub}{\Dz\mun\neumb} channel, (top right) the \decay{\Bub}{\Dstarz\mun\neumb} channel, and (bottom) the \decay{\Bub}{\dstst\mun\neumb} channel. The contributions from events in which there is only one physical solution, in which there are two and the chosen lower energy solution is correct, or in which the incorrect solution is chosen are shown.\label{fig:mmsq_resp}}
\end{figure}

\begin{figure}[tbp]
  \begin{center}
    \begin{subfigure}{0.7\textwidth}
      \includegraphics[width=\textwidth]{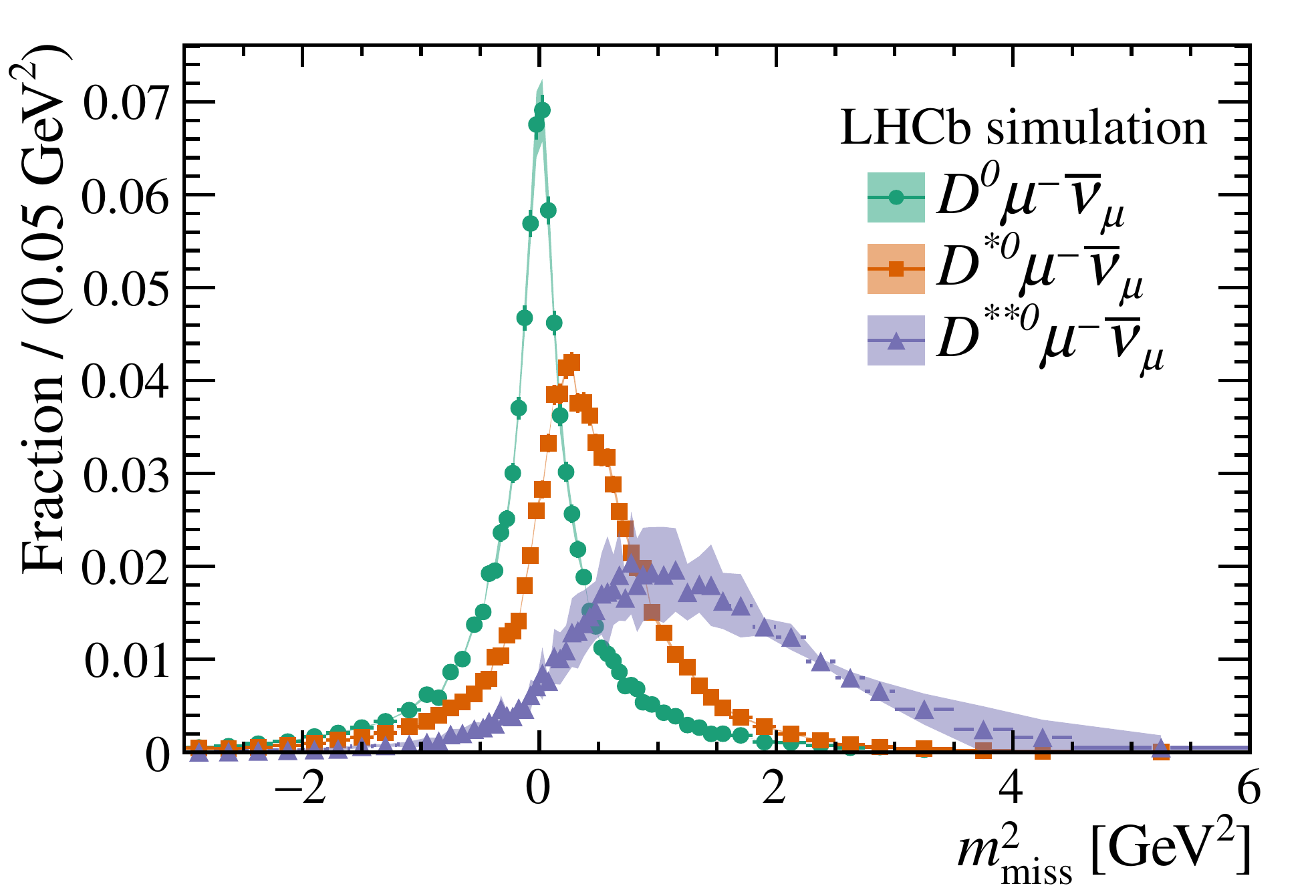}
    \end{subfigure}
  \end{center}
  \caption{ The missing mass shapes from simulation for the signal samples are shown. The bands around the points represent the systematic uncertainties on the form factors in the simulation and the branching fractions for different contributions to the \dstst channel.\label{fig:mmsq_sig}}
\end{figure}

We keep only the physical solutions for $E_B$ which are greater than the sum of the energies of the reconstructed decay products.
Based on simulation, approximately 75\% of signal candidates have a physical solution.  For candidates with two physical solutions, the one with lower energy is correct 90\% of the time. Only the lower energy solution is used for these candidates.
The difference $\Delta\mmsq$ between the reconstructed missing-mass squared and the corresponding true values for different classes of solutions are shown in \cref{fig:mmsq_resp}. When $E_B$ is correctly reconstructed, the full-width at half maximum of the $\Delta\mmsq$ distribution is approximately \SI{0.4}{\giga\evolt\squared} and is consistent among the signal channels.
The resulting \mmsq distributions for the signal decays to be used in the fit are shown in \cref{fig:mmsq_sig}.

\section{Background estimation}
\label{sec:backgrounds}

The backgrounds to the \bstwostb signal candidates come from a number of different sources. For each of these sources, we estimate the overall yield as well as the missing-mass shapes. The most important sources are semileptonic decays of \Bub and \Bdb mesons not originating from a \bstwostb or \bsoneb decay, which represent 83\% of the total number of selected candidates.

\begin{figure}[tb]
  \begin{subfigure}{0.5\textwidth}
    \includegraphics[width=\textwidth]{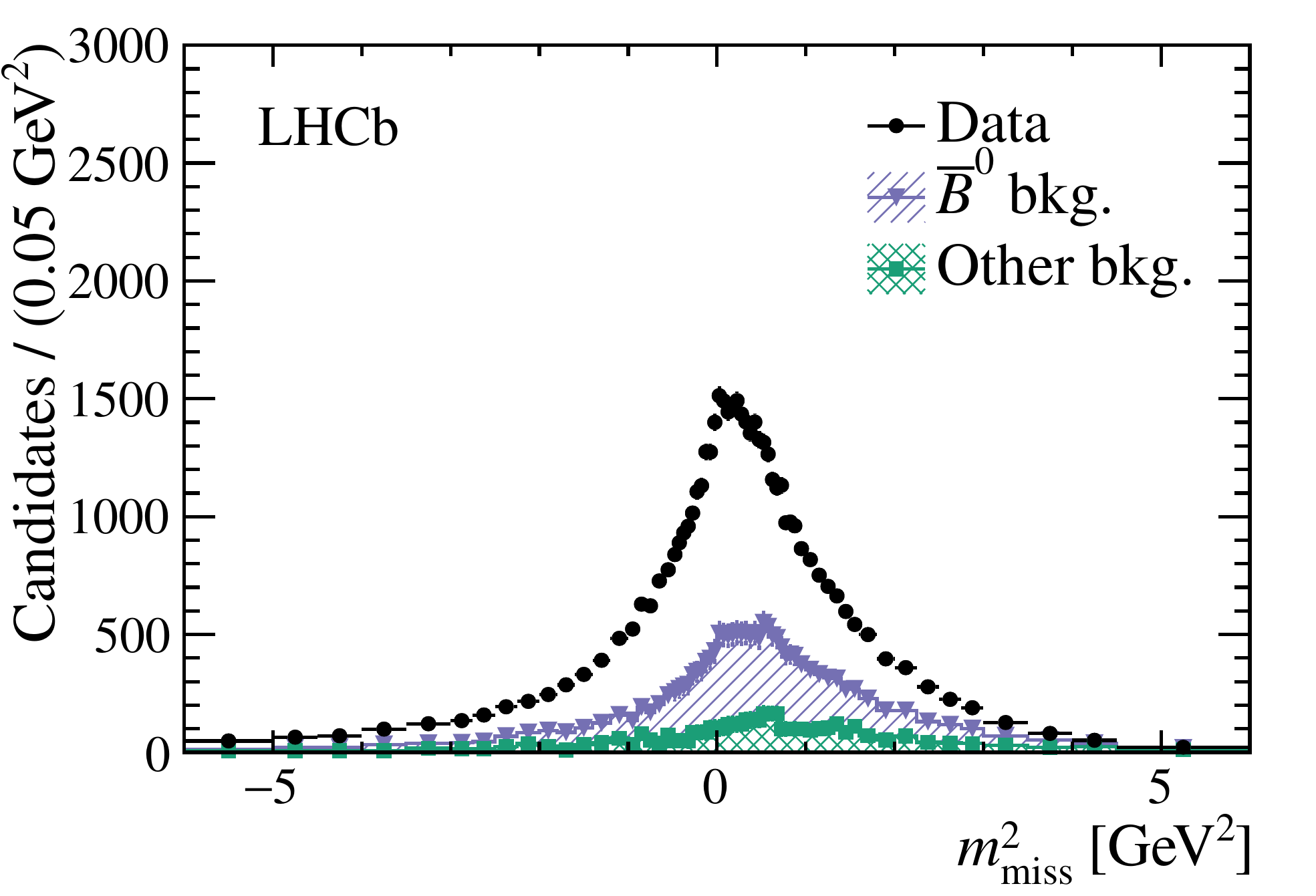}
  \end{subfigure}
  \begin{subfigure}{0.5\textwidth}
    \includegraphics[width=\textwidth]{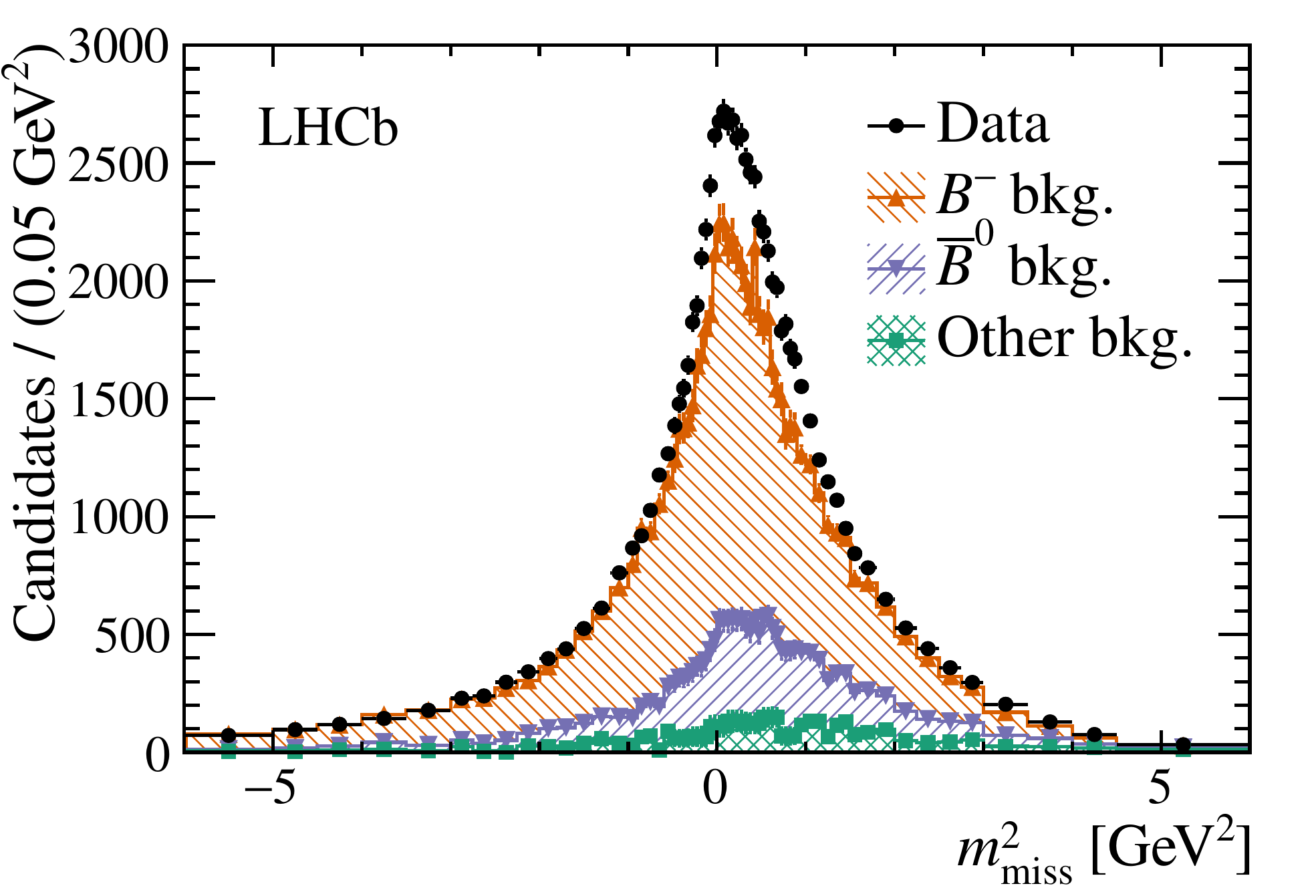}
  \end{subfigure}
  \caption{Missing-mass distribution for data and estimated background contributions in the (left) same-sign kaon sample and (right) opposite-sign sample. The other background decays include contributions from misreconstructed backgrounds, and semileptonic decays of \Bsb and \Lb mesons. The remainder of the \SSK sample not from \Bdb or other background decays is used to define the background contribution from \Bub semileptonic decays. This is then extrapolated to the \OSK sample, where the remainder is composed of signal. The background distributions are stacked.\label{fig:mmsq}}
\end{figure}

The overall estimated background in the \mmsq distribution is shown in \cref{fig:mmsq}. We make this estimation by first considering a number of smaller contributions not from semileptonic decays of \Bub and \Bdb mesons:
  \begin{itemize}
  \item misreconstructed backgrounds consisting of
  \begin{itemize}
  \item non-\Dz backgrounds,
  \item $\Dz\mun$ combinations not from the same $b$-hadron decay,
  \item backgrounds with a hadron misidentified as the muon;
  \end{itemize}
  \item \Bsb and \Lb semileptonic decays to final states including a \Dz meson.
\end{itemize}
Together, these backgrounds total 8\% of all selected candidates. We estimate their yield and shape in both the \mmsq and the \dmmin variables as described in \cref{subsec:smallbkg}. These can then be accounted for in both the distributions of the \OSK and \SSK data samples.
We then estimate the semileptonic \Bub and \Bdb backgrounds as described in \cref{subsec:slbkg}. The expectation for the \Bdb contribution is subtracted from the remaining \SSK sample, producing an estimate for the shape of the \Bub contribution in that sample. These two distributions are then extrapolated to the \OSK sample to produce the background estimation. The difference between this estimation and the full \OSK yield is composed of signal decays.

\subsection{Backgrounds not from semileptonic decays of $\boldsymbol{\Bub}$ and $\boldsymbol{\Bdb}$ mesons}
\label{subsec:smallbkg}

Misreconstructed backgrounds are estimated using data-driven techniques. The yields and \dmmin and \mmsq shapes of backgrounds without a \Dz meson are estimated using sidebands around the \Dz mass peak. The sideband ranges chosen are from \SIrange{1790}{1830}{\mega\evolt} and from \SIrange{1900}{1940}{\mega\evolt}. The difference of the \mmsq shape between the left and right sidebands is negligible. Approximately 3\% of the selected candidates come from this background.

Combinations of $\Dz\mun$ not coming from a single $b$-hadron decay are estimated using a wrong-sign ($\Dz\mup$) control sample, assuming that the doubly Cabbibo-suppressed contribution from \decay{\Dz}{\Kp\pim} is negligible.  Along with this estimation, the contributions from misidentified muons to both the signal and wrong-sign samples are estimated using a control sample with particle-identification requirements that remove true muons.  We then weight this sample using-particle identification efficiencies derived from calibration samples~\cite{LHCb-PUB-2016-021} to estimate the misidentified muon contamination. Together these two sources make up less than 1\% of selected candidates.

We use a combination of data and simulation to estimate backgrounds from \decay{\Bsb}{\Dz K^{+} X \mun \neumb}, \decay{\Bsb}{\Dz K^{0} X \mun \neumb}, and \decay{\Lb}{ \Dz p X \mun \neumb} decays.  In data, additional candidates identified as kaons or protons, which are inconsistent with being produced at any primary collision vertex, are combined with the $\Dz\mun$ candidates.  This is done for both right- ($\Dz\Kp$ or $\Dz p$) and wrong-sign ($\Dz\Km$ or $\Dz \antiproton$) combinations.  The wrong-sign combinations are used to model the combinatorial background in this selection.  Using a-two dimensional fit to the $\Dz K$ or $\Dz p$ mass and the track impact parameter with respect to the $\Dz\mun$ vertex, we determine the \Bsb and \Lb yields.

For the \Bsb case, the resulting yield is corrected for efficiency, and for modes with neutral kaons, using simulation. We take the shape of the contribution in \dmmin from simulation. There is an important contribution at low \dmmin where the kaon from the \Bsb decay points back to the primary vertex and is selected as the prompt kaon.  This contribution is not present in the data control sample because of the requirement for the additional kaon to be inconsistent with any primary vertex.  The final cut on \dmmin does, however, remove this component from the signal region. The simulated samples well reproduce the shape of the \dmmin distribution measured using the $\Dz\Kp X\mun$ selection.

\begin{figure}[tb]
  \begin{subfigure}{0.5\textwidth}
    \includegraphics[width=\textwidth]{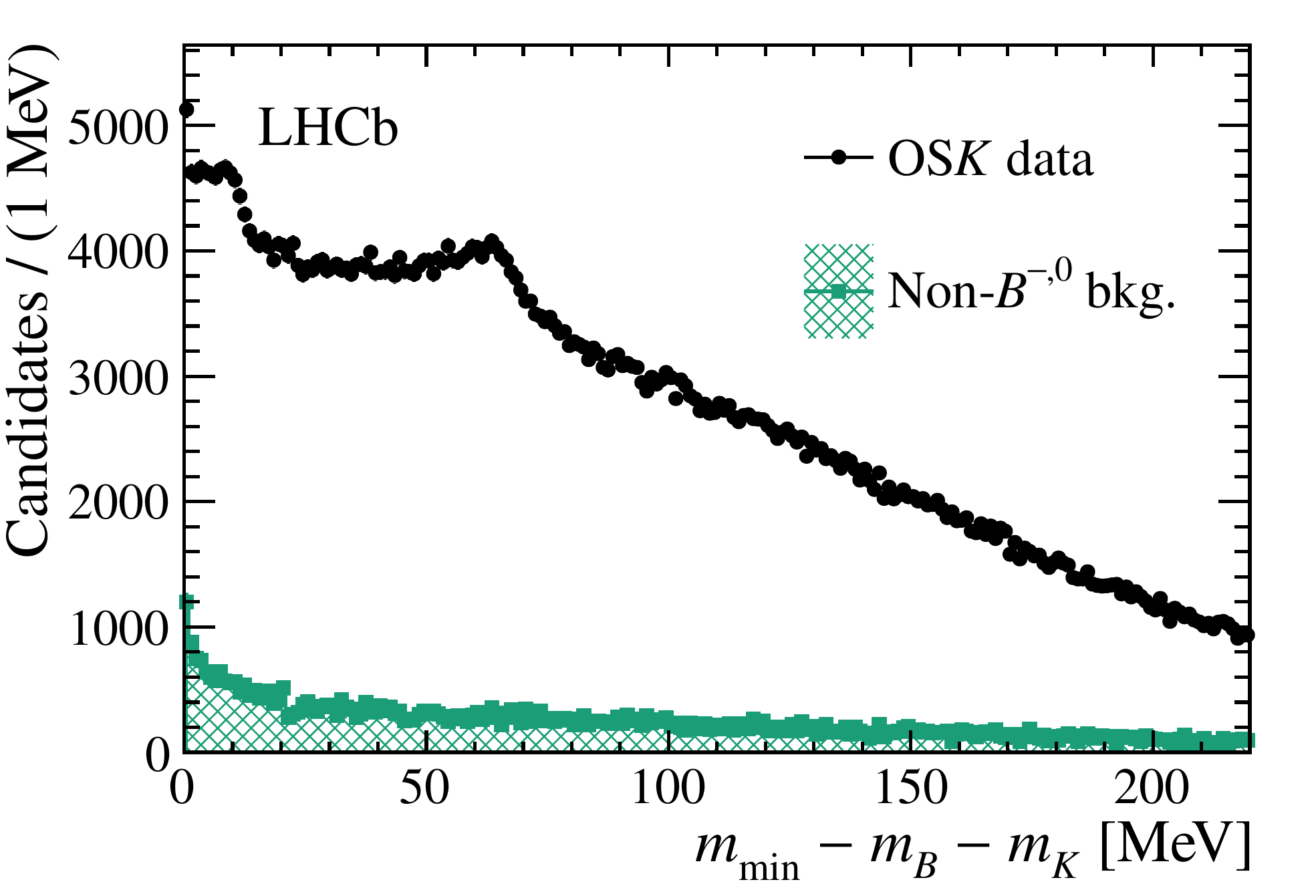}
  \end{subfigure}
  \begin{subfigure}{0.5\textwidth}
    \includegraphics[width=\textwidth]{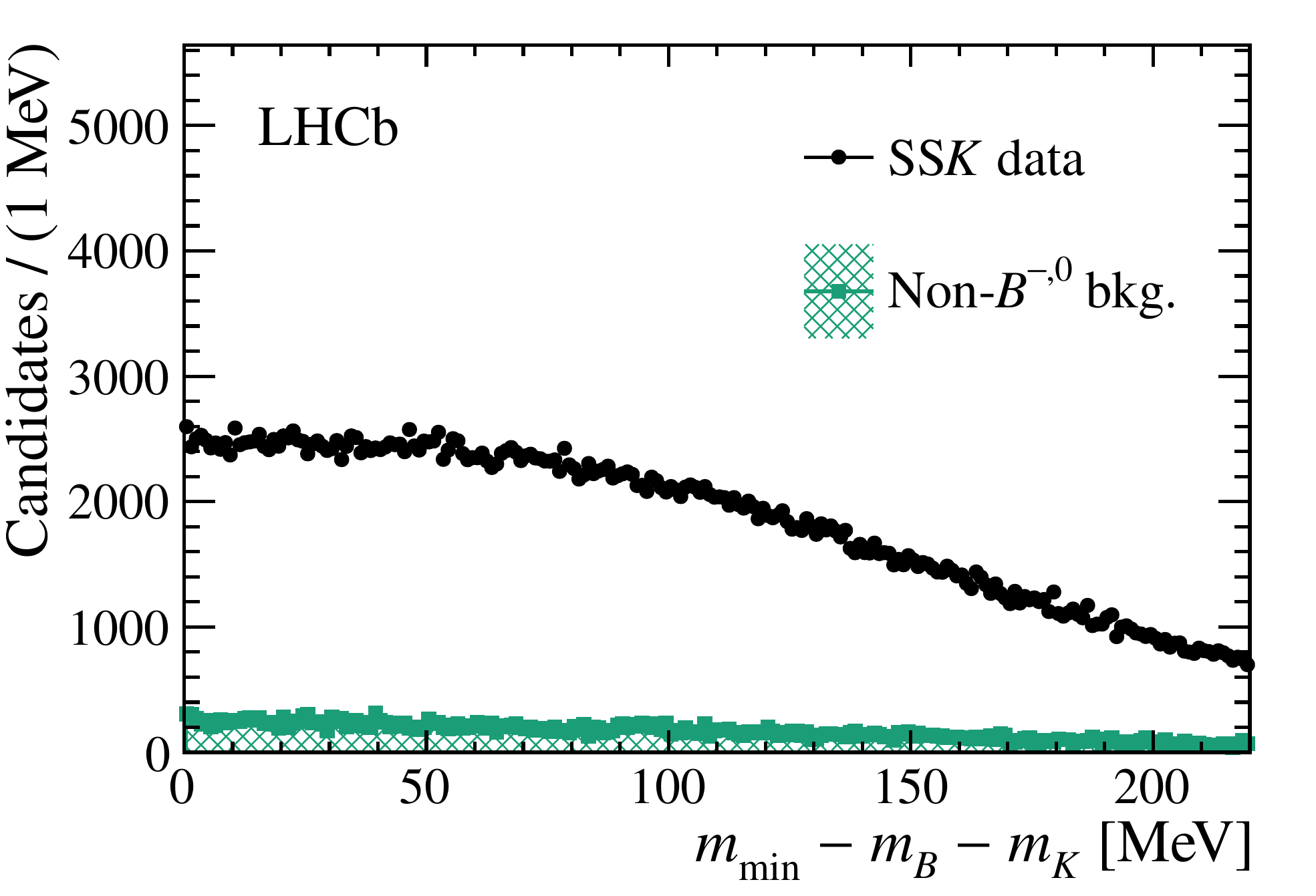}
  \end{subfigure}
  \caption{Distribution of the minimum mass difference for (left) $\Bub\Kp$ opposite-sign candidates and (right) $\Bub\Km$ same-sign candidates. All candidates are compared to the estimated background from other sources besides decays of a \Bub or \Bdb meson to $\Dz X\mun\neumb$. The remaining non-peaking part of the distributions is made up of \Bub and \Bdb semileptonic decays that do not come from an excited \Bsb state.\label{fig:mmin_wbkg}}
\end{figure}

Since the simulation does not reproduce well the shape in \mmsq for the $\Dz\Kp X\mun$ control sample, the shape of the \Bsb contribution to the main \mmsq fit is instead derived from the control sample. We obtain it by taking the difference in the right- and wrong-sign kaon \mmsq distributions, scaling the wrong-sign yield to match the combinatorial contribution found by the two dimensional fit described above. The \Bsb contribution to the final selection is 3\%, with a relative normalization uncertainty of 10\%. For the \Lb case, the contribution is less than 1\%. The shapes in both \dmmin and \mmsq are taken from the control sample, and scaled based on the efficiency in simulation.  The relative uncertainty on the normalization of this contribution is 20\%.
The \dmmin distribution for the sum of these backgrounds is shown in \cref{fig:mmin_wbkg}.

\subsection{Backgrounds from semileptonic decays of $\boldsymbol{\Bub}$ and $\boldsymbol{\Bdb}$ mesons}
\label{subsec:slbkg}

We first estimate the number of candidates in the \OSK signal region that do not come from \bstwostb decays.  This is done with a fit to the \dmmin distribution in the control region after subtracting the backgrounds described in \cref{subsec:smallbkg}. The fit is done for three bins of prompt kaon \pt to account for the different spectra of the \SSK and \OSK samples: $0.5 < \pt < \SI{1.25}{\giga\evolt}$, $1.25 < \pt < \SI{2}{\giga\evolt}$, and $\pt > \SI{2}{\giga\evolt}$.
The \dmmin shapes for \decay{\bstwostb}{\Bub\Kp} signals as well as \bsoneb and \decay{\bstwostb}{\B^{*-}\Kp}, with \decay{\B^{*-}}{\Bub\gamma}, backgrounds are taken from simulation.  We model the background contribution using a fifth-order polynomial; the high order allows the fit to account for additional backgrounds peaking near $\dmmin = 0$.

\begin{figure}[tb]
  \begin{subfigure}{0.5\textwidth}
    \includegraphics[width=\textwidth]{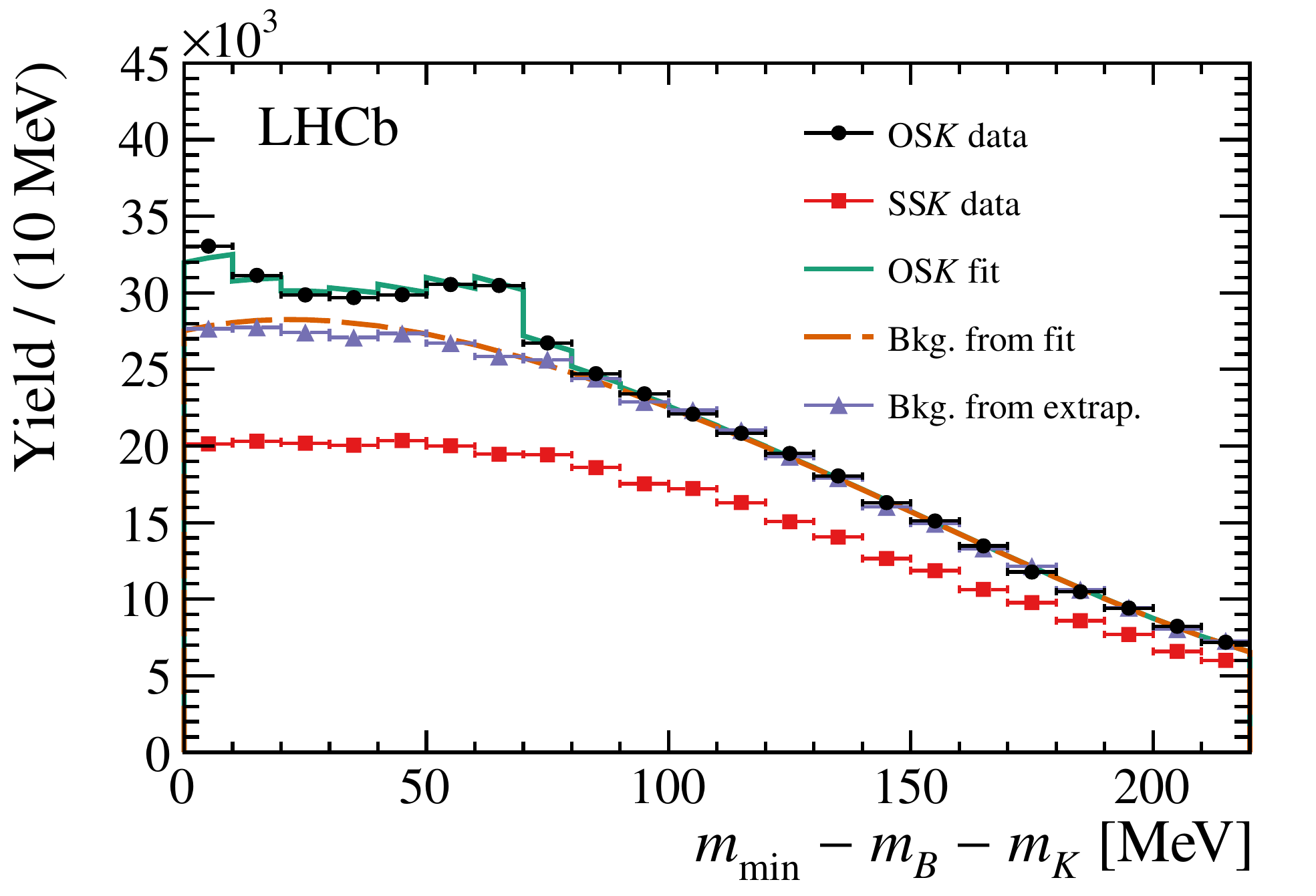}
  \end{subfigure}
  \begin{subfigure}{0.5\textwidth}
    \includegraphics[width=\textwidth]{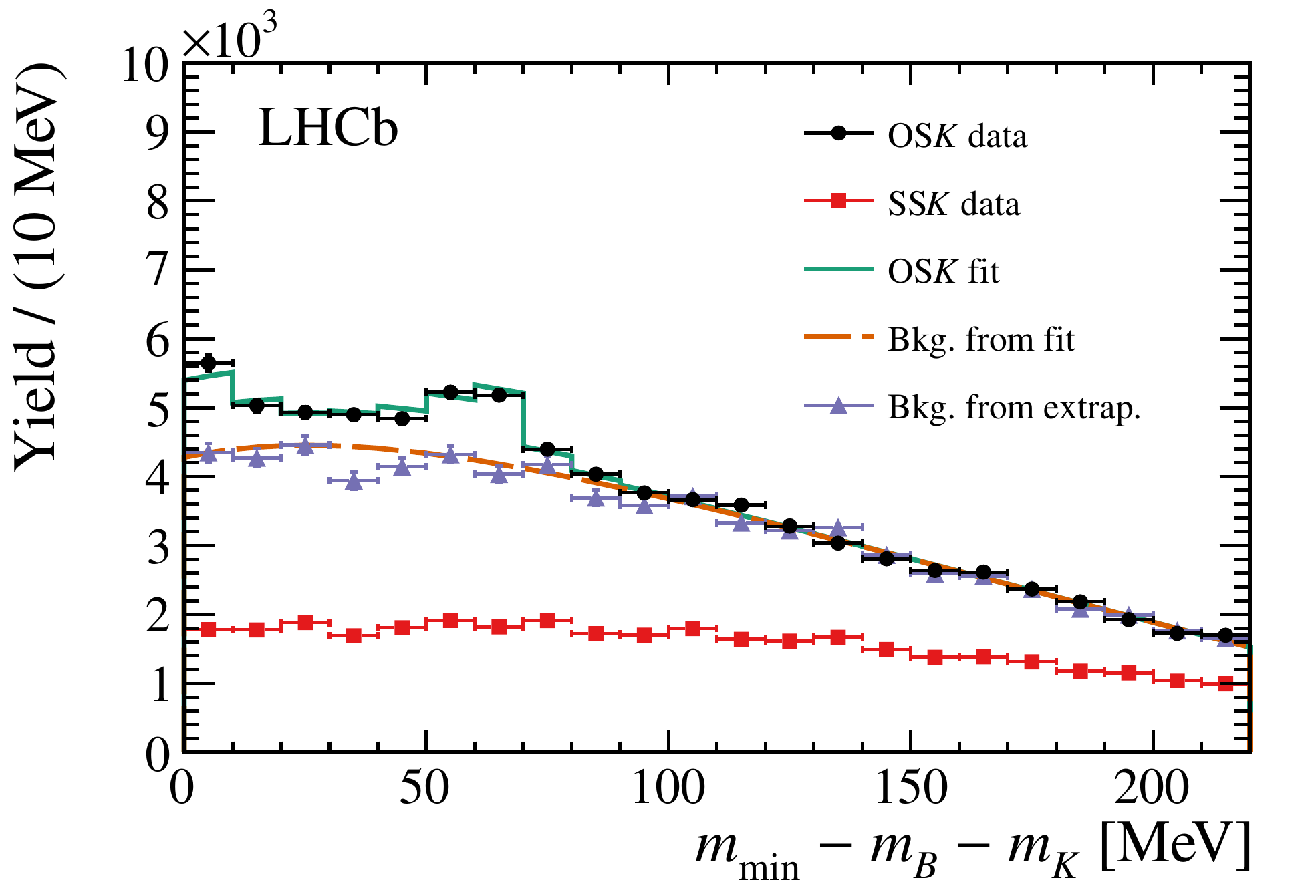}
  \end{subfigure} \\
  \begin{center}
  \begin{subfigure}{0.5\textwidth}
    \includegraphics[width=\textwidth]{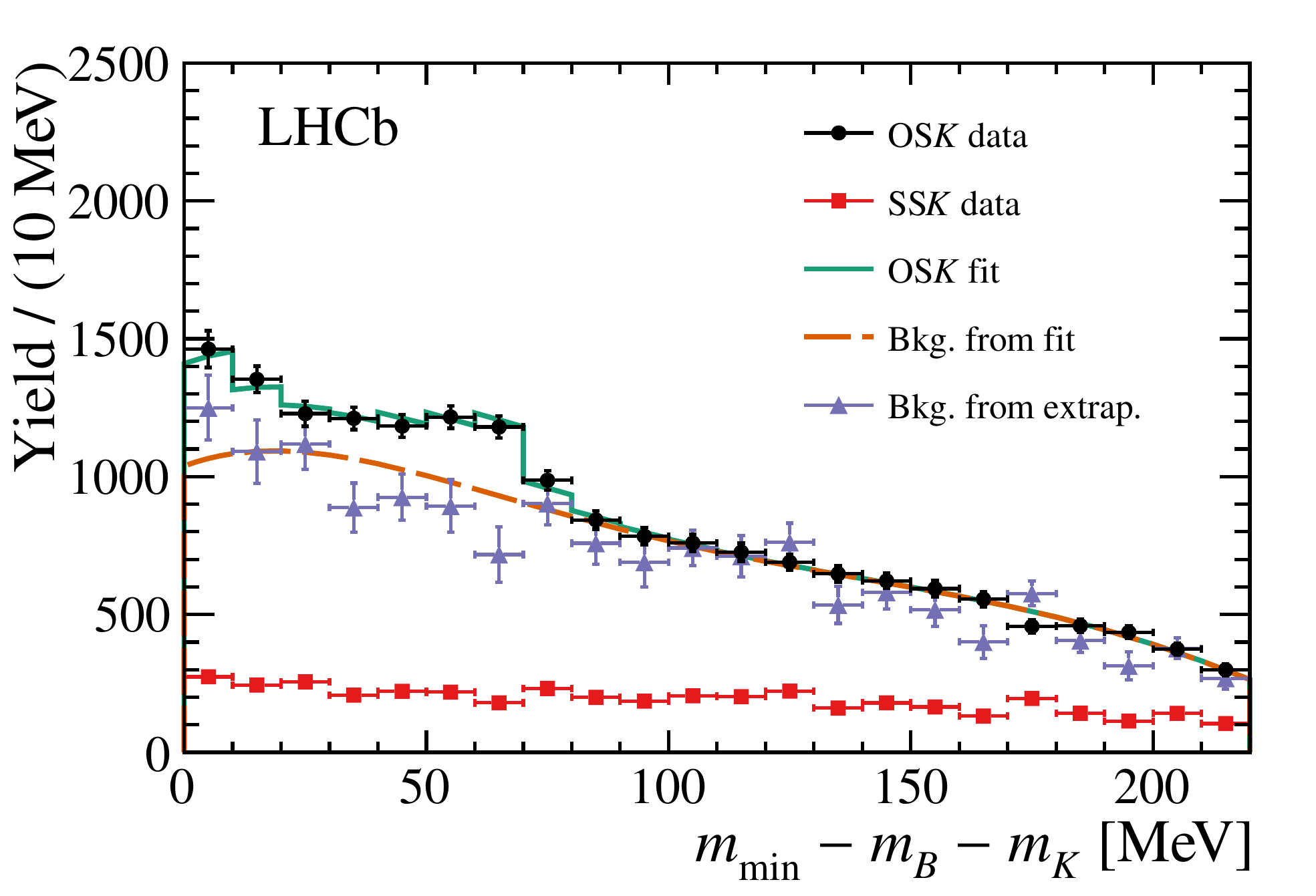}
  \end{subfigure}
  \end{center}
  \caption{Fits to the opposite-sign and same-sign kaon $\mmin - m_B - m_K$ distributions with non-\Bub and \Bdb backgrounds subtracted, and the resulting estimations of the non-\bstwostb and \bsoneb contributions. The fits are done separately in three bins of the prompt kaon \pt: (top left) $0.5 < \pt < 1.25\gev$, (top right) $1.25 < \pt < 2\gev$, and (bottom) $\pt > 2\gev$. The dashed line shows the background estimation using a fit to the full \OSK distribution with signal templates from simulation and a fifth-order polynomial for the background. The points estimate the background using a linear extrapolation of the \OSK to \SSK ratio in the region $\mmin - m_B - m_K > 100\mev$.\label{fig:mmin_fit}}
\end{figure}

In an alternative approach, the \SSK sample is scaled to model the background in the \OSK sample. The scaling is based on a linear fit to the ratio between \OSK and \SSK samples in the region $\dmmin > \SI{100}{\mega\evolt}$, where the signal contribution is negligible. The \dmmin distributions, showing the results of these two methods of background estimation, are shown in \cref{fig:mmin_fit}. We use the difference of the two methods to estimate the systematic uncertainty on the background yield.

The two methods constrain the yield of non-\bstwostb decays as a function of \dmmin, however the missing-mass shape in the \OSK channel must still be determined. For each type of background decay, the missing-mass distribution is the same in the \OSK and \SSK samples for a particular value of \dmmin. This equivalence is tested using fully reconstructed \decay{\Bub}{\jpsi \Km} decays. However, since the missing mass also depends on the decay products, the distributions are different for \Bub and \Bdb decays. The fraction of this background coming from \Bdb decays is also different in the \SSK and \OSK samples.

We use the \SSK shape to model the background contribution in the \OSK sample, considering \Bub and \Bdb  decays separately. This is done by estimating first the contribution of \Bdb decays to both the \OSK and \SSK channels. The remainder of the \SSK channel is used to model the shape of the \Bub contribution. The normalization of the \Bub background in the \OSK channel is then derived from the overall non-\bstwostb contribution with that from \Bdb mesons removed.

To estimate the fractional contribution from \Bdb decays in \SSK sample, we use the expected fraction resulting in the final state $\Dz X\mun \neumb$ based on measured branching fractions~\cite{mynote}.  The overlap with this measurement is removed by considering separately the ratio of contributions to the final state from \Bdb and \Bub decays for the \decay{\B}{D^{*(*)}\mun\neumb} channels, $r_{D^{*(*)}}$, with \decay{D^{*(*)}}{\Dz X}.
These ratios are combined with the measured fractions $f_{\Dstarz}$ and $f_{\dstst}$. We assume equal production of \Bdb and \Bub mesons.  The fraction of \Bdb decays in the \SSK sample, $f_{\Bdb}$, is thus given by
\begin{align}
    \frac{1}{f_{\Bdb}} &= \frac{ \BF\qty( \decay{\Bdb}{\Dz X\mun\neumb}) + \BF\qty( \decay{\Bub}{\Dz X\mun\neumb})}{  \BF\qty( \decay{\Bdb}{\Dz X\mun\neumb}) } \nonumber \\
  &= 1 + \qty[ \frac{ \BF\qty( \decay{\Bdb}{\Dstarp\mun\neumb})\BF\qty(\decay{\Dstarp}{\Dz X}) + \BF\qty( \decay{\Bdb}{D^{**+}\mun\neumb})\BF\qty( \decay{D^{**+}}{\Dz X} )  }{ \BF\qty( \decay{\Bub}{\Dz X\mun\neumb}) } ]^{-1} \nonumber \\
   &= 1 + \qty[ r_{\Dstar} f_{D^{*0}} + r_{D^{**}}f_{D^{**0}} ]^{-1} \nonumber \\
  &=  1 + \qty[ (0.591\pm 0.024) f_{D^{*0}} + (1.00 \pm 0.23)f_{D^{**0}} ]^{-1}.
\end{align}
The uncertainty on $r_{\Dstar}$ comes chiefly from experimental uncertainty, while the dominant uncertainty on $r_{D^{**}}$ comes from extrapolation to the unmeasured parts of the semileptonic width. The uncertainty is taken as one standard deviation of the full extrapolation envelope assuming a uniform distribution.  Using the central values of the expectations for $f_{\Dstarz}$ and $f_{\dstst}$ given in \cref{sec:intro}, the central value for $f_\Bdb$ is 35\%; variations within the uncertainties change it by approximately 2\%. We then combine this value of $f_\Bdb$ with an efficiency correction from simulation which depends on the lifetime difference between \Bub and \Bdb mesons.

The contribution from \Bdb mesons is studied similarly to the \Bsb and \Lb backgrounds, by attaching an additional candidate identified as a pion to the $\Dz\mun$ candidates.  We fit the $\Dz\pi^{\pm}$ mass distributions, including peaking contributions from \Dstarp, $D_1$, and $D_2^*$ mesons on top of a smooth distribution.  The normalizations of the peaks from the decay \decay{\Bdb}{ \qty( \decay{D_2^{*+}}{\Dz \pip})\mun\neumb} and the partially reconstructed decays \decay{\Bdb}{ \qty( \decay{ D_1^+  }{ \Dstarz \pip} )\mun\neumb} and \decay{\Bdb}{ \qty( \decay{ D_2^{*+}  }{ \Dstarz \pip} )\mun\neumb} show that there are more  \Bdb candidates in the \OSK sample than there are in the \SSK sample. This is verified using fully reconstructed \decay{\Bdb}{\jpsi \Kstarb \qty(892)^0} decays. Combining the ratios in the two channels, we find there is a 10\% larger contribution of \Bdb decays in the \OSK sample.

While the decays in the resonance peaks are dominated by either a \Bub or \Bdb initial state, the other contributions to the $\Dz\pi^{\pm}$ distributions are more difficult to disentangle. The combinatorial background is expected to be symmetric in $\Dz\pip\mun$ and $\Dz\pim\mun$, while \Bub decays produce $\Dz\pip\pim\mun$ which also contribute equally to both distributions.  We therefore derive the \Bdb missing-mass shape by subtracting the $\Dz\pim\mun$ shape from the $\Dz\pip\mun$ shape. Each shape is corrected for the efficiency to reconstruct the additional pion based on simulation. The resulting distribution is validated using a simulated mixture of \Bdb decays.

We determine the total background shape from \Bub and \Bdb decays in the \OSK sample by first removing the expected \Bdb contribution from the initial \SSK sample's \mmsq distribution. This is then scaled up by 10\% to estimate the \Bdb contribution to the \OSK sample.  The remainder of the \SSK sample, composed of \Bub decays, is scaled up so that when it is added to the \Bdb estimate, the total number of background candidates in the \OSK sample is equal to the result of the \dmmin fit.  We accomplish this procedure using an event-by-event weighting that accounts for the background yield as a function of \dmmin.

Contributions not from semileptonic decays of \Bub and \Bdb mesons that are subtracted from the \SSK sample (\Bsb and \Lb contributions, combinatorial, and misidentified muons) are also weighted in the same manner before being subtracted to produce the final background template.

\subsection{Backgrounds from $\boldsymbol{\bstwostb}$ and $\boldsymbol{\bsoneb}$ decays}

The final class of backgrounds are \bsststb decays that produce a \Bub meson with a $\Dz\mun X$ final state that is not a semileptonic channel of interest. The \mmsq shapes for semitauonic \decay{\Bub}{ \Dz X \qty(\decay{\taum}{\mun \neumb \neut})\neutb} decays and \Bub decays involving two charm mesons are estimated from simulation, and are included in the final fit. Contributions from \bsoneb or \decay{\bstwostb}{B^{*-}\Kp}, where \decay{B^{*-}}{\Bub\gamma}, are negligible after the requirement on the \dmmin variable.

\section{Fit description}
\label{sec:fit}

The fractions of interest, $f_\Dz$ and $f_\dstst$, are determined from a binned-template, maximum-likelihood fit to the missing-mass distribution of the \OSK sample. The signal fraction $f_{\Dstarz}$ is given by the remainder, $1 - f_{\Dz} - f_\dstst$. To control statistical fluctuations in the templates for the missing-mass tails, which are important for determining the \dstst content, a variable bin size is used for the template fit. The sum of the templates is allowed to vary bin-by-bin based on the combined statistical uncertainty of all templates.
This variation is included using a single nuisance parameter for each bin that is constrained by the statistical uncertainty.
It is dominated by the uncertainty of the \SSK sample used to create the combined \Bub and \Bdb background template. The effect of these uncertainty parameters is determined analytically using the Barlow--Beeston method~\cite{Barlow:1993dm}.
Unless otherwise specified, we account for systematic uncertainties using nuisance parameters that are free to vary in the fit; these parameters are allowed to vary around their central values with a Gaussian constraint based on their uncertainty.

In total, the fit contains three signal and eight background templates: background from semileptonic \Bub and \Bdb decays not from a \bstwostb decay, non-\Dz backgrounds, $\Dz\mun$ combinations not from the same $b$-hadron decay, backgrounds with a hadron misidentified as the muon, \Bs, \Lb, \bstwostb decays with a semitauonic \Bub decay, and \bstwostb decays with a \Bub decay to two charm mesons. There are 18 free parameters in the fit, not including the nuisance parameters for the template statistical uncertainties.

The three templates describing the signal are obtained from simulation---exclusive \Dz, exclusive \Dstarz, and the sum of all \dstst modes; these are shown in \cref{fig:mmsq_sig}. We also correct for the relative reconstruction and selection efficiencies between these samples, which are taken from simulation. Relative to the \Dstarz mode, the efficiency of the \Dz mode is 92\% and that of the \dstst mode is 68\%. In addition to the two signal fractions of interest, three more free parameters govern the shape changes from the variations of the form factors, and one parameter gives the overall signal yield.

The template describing the \Bub and \Bdb backgrounds not coming from a \bstwostb meson is extrapolated from the \SSK sample as described in \cref{sec:backgrounds}. Four free parameters describe the systematic variations of the normalization as a function of \dmmin.
In the fit, the parameters $r_{\Dstar}$ and $r_{D^{**}}$ and the fractions $f_\Dz$ and $f_\dstst$ are used to calculate $f_{\Bdb}$ for the current evaluation of the fit function. This variation is constrained by the uncertainties of $r_{\Dstar}$ and $r_{D^{**}}$.  The current value of $f_\Bdb$ is combined with a set of templates that vary $f_{\Bdb}$ by ${}\pm1\%$ to extrapolate from the nominal value and produce the estimated background shape for this evaluation. An additional uncertainty in this template comes from the \mmsq shape of the \Bdb component, which is controlled by one parameter.

The normalizations of the contributions from \Bsb decays, \Lb decays, and decays involving misidentified muons are also allowed to vary. The data-driven background shapes for fake and combinatorial muons, and for \Bsb and \Lb decays are described in \cref{sec:backgrounds}.

The templates for the contribution of semitauonic decays of \Bub mesons from \bstwostb are obtained from simulation.  We determine the normalization relative to the semimuonic modes by deriving an effective ratio of semitauonic to semimuonic decays, $R\qty(\Dz X)$, using the Standard Model values~\cite{Na:2015kha,Bigi:2017jbd,Bernlochner:2016bci} and the expected fractions of \Dz, \Dstarz, and \dstst,
\begin{equation}
R\qty( \Dz X) = R\qty(D)f_\Dz + R\qty(D^*)f_\Dstarz + R\qty(D^{**})f_\dstst,
\end{equation}
where $R\qty(D)$ is the ratio $\BF\qty(\decay{\Bb}{D\taum\neutb})/\BF\qty(\decay{\Bb}{D\mun\neumb})$, and $R(D^*)$ and $R(D^{**})$ are the corresponding ratios in the other decay channels. This is combined with the $\tau\to\mu X$ branching fraction~\cite{PDG2016} and the relative efficiency to reconstruct $\tau$ decays taken from simulation.  The expected contribution is $\qty( 1.5 \pm 0.3 )\%$ of the selected \bstwostb decays. The uncertainty is dominated by the difference of the Standard Model expectations and the world-average measured values of $R(D)$ and $R(D^*)$~\cite{HFLAV16}, which we take as a systematic uncertainty.

The other backgrounds coming from \decay{\bstwostb}{\Bub\Kp} decays are \Bub mesons decaying to double-charm states of various types. A simulated sample composed of many different decays producing $\Dz\mun$ final states is used to determine the shape of this component.  The normalization of the resulting missing-mass template is expected to be about 1\% of \bstwostb decays based on branching fractions, but is left unconstrained in the fit.

\section{Systematic uncertainties}
\label{sec:syst}

Each of the signal components has  systematic uncertainties associated to its shape.  The systematic uncertainty on the \Dz and \Dstarz components is estimated based on uncertainties in the form-factor parameters. We reweight our simulated samples using the Caprini--Lellouch--Neubert (CLN) expansion formalism~\cite{Caprini:1997mu}, with the uncertainties on the parameters taken from HFLAV~\cite{HFLAV16}. This produces negligible changes in the missing mass template shapes compared to the other uncertainties in this analysis.

The uncertainty on the relative signal efficiencies is approximately 2\%. We obtain the associated systematic uncertainty by repeating the fit with different efficiency values obtained by varying the efficiencies by their uncertainties.

For the \dstst template, in addition to a large variation in the form-factor distribution based on results from Ref.~\cite{Bernlochner:2016bci}, we create an alternative template with different branching fractions for the various resonant and nonresonant decay modes. The most important difference is the inclusion of a larger fraction of higher mass, nonresonant $D^{(*)}\pi$ and $D^{(*)}\pi\pi$ decays, where the pions may be of any allowed charge combination. This shape is fixed in the template fit; a second fit with the alternative template is used to estimate the systematic uncertainty from this shape. During this second fit, the signal efficiency of the \dstst component is also adjusted along with the template. This uncertainty leads to the bands shown in \cref{fig:mmsq_sig}.

For background contributions not from \Bub or \Bdb semileptonic decays, we include individual uncertainties on their normalizations. Systematic variations in the shapes are dominated by the statistical bin-by-bin statistical uncertainty.

We consider a number of systematic uncertainties on the \Bub and \Bdb contributions. The uncertainty due to the overall normalization comes from two sources. The statistical uncertainties in the polynomial background function of the \dmmin fit are used to modify the template.  This corresponds to an uncertainty of less than 1\% on the yield in each prompt kaon \pt bin. We also use the alternative extrapolation using the \dmmin ratio to provide an alternative normalization, giving an uncertainty of approximately 2\%. Both of these uncertainties produce only small changes in the templates.
The uncertainties in $r_{\Dstar}$ and $r_{D^{**}}$ give the uncertainty on the \Bdb fraction. The uncertainty in the \Bdb \mmsq shape is estimated from the uncertainty in the efficiency from simulation to reconstruct the pion in the $\Dz\pi^{\pm}\mun$ combination.

An estimated breakdown of the total statistical and systematic uncertainty is given in \cref{tab:uncertainty}. The largest source of uncertainty is the statistical uncertainty from the extrapolated \SSK data sample. The uncertainty in the \Bdb \mmsq shape is also important because of its effect on the high \mmsq tail.
Most systematic uncertainties are included in the fit with constrained nuisance parameters.  The only source for which the fit result has a significantly smaller uncertainty than the initial constraint is the normalization of the non-\bstwostb 
background from the \dmmin extrapolation. 
For the final result, the total uncertainty is taken from the best fit, with the fixed systematic uncertainties for the relative signal efficiencies and the \dstst branching fractions from added in quadrature.

\begin{table}[tb]
  \caption{Estimates of the breakdown of the total uncertainty. All estimates are done by repeating the fit with systematic nuisance parameters fixed to their best fit values. The statistical uncertainty of the \OSK sample is estimated from the uncertainty on the signal fractions with the template statistical nuisance parameters fixed to their best fit values. The template statistical uncertainty is added in by allowing only the statistical nuisance parameters to vary. The effect of each floating systematic uncertainty is estimated by refitting with its systematic nuisance parameter shifted by the uncertainty found by the best fit and taking the difference in the signal fractions as the uncertainty. The total uncertainty is taken from the best fit, with the fixed systematic uncertainties added in quadrature.\label{tab:uncertainty}}
  \begin{center}
  \begin{tabular}{llS[table-format = 1.3]@{\extracolsep{1cm}}S[table-format = 1.3,retain-explicit-plus]}
    \toprule
    \multicolumn{2}{c}{Source of uncertainty} & \multicolumn{1}{c}{$f_\Dz$} & \multicolumn{1}{c}{$f_\dstst$} \\
    \midrule
    \multirow{2}{*}{Statistical} & \OSK sample & 0.025  & 0.027 \\
    & Templates & 0.047 & 0.052 \\
    \midrule
    \multirow{3}{*}{Floating syst.} & Signal form-factors & 0.006 & 0.004 \\
    & Non-\Bub, \Bdb backgrounds & 0.004 & 0.004 \\
    & \Bub, \Bdb background normalization & 0.003  & 0.015 \\
    & \Bdb fraction and \mmsq shape & 0.004 & 0.030 \\
    \midrule
    \multirow{2}{*}{Fixed syst.} & \dstst branching fractions & 0.025 & 0.044 \\
    & Relative signal efficiency & 0.003 & 0.003 \\
    \midrule 
    \multicolumn{2}{l}{\multirow{2}{*}{Total uncertainty }} & {\multirow{2}{*}{0.056}} & +0.070 \\
    & & & -0.074 \\
    \bottomrule
  \end{tabular}
  \end{center}
\end{table}

\section{Results and conclusions}
\label{sec:results}

\begin{figure}[htbp]
  \begin{subfigure}{0.5\textwidth}
    \includegraphics[width=\textwidth]{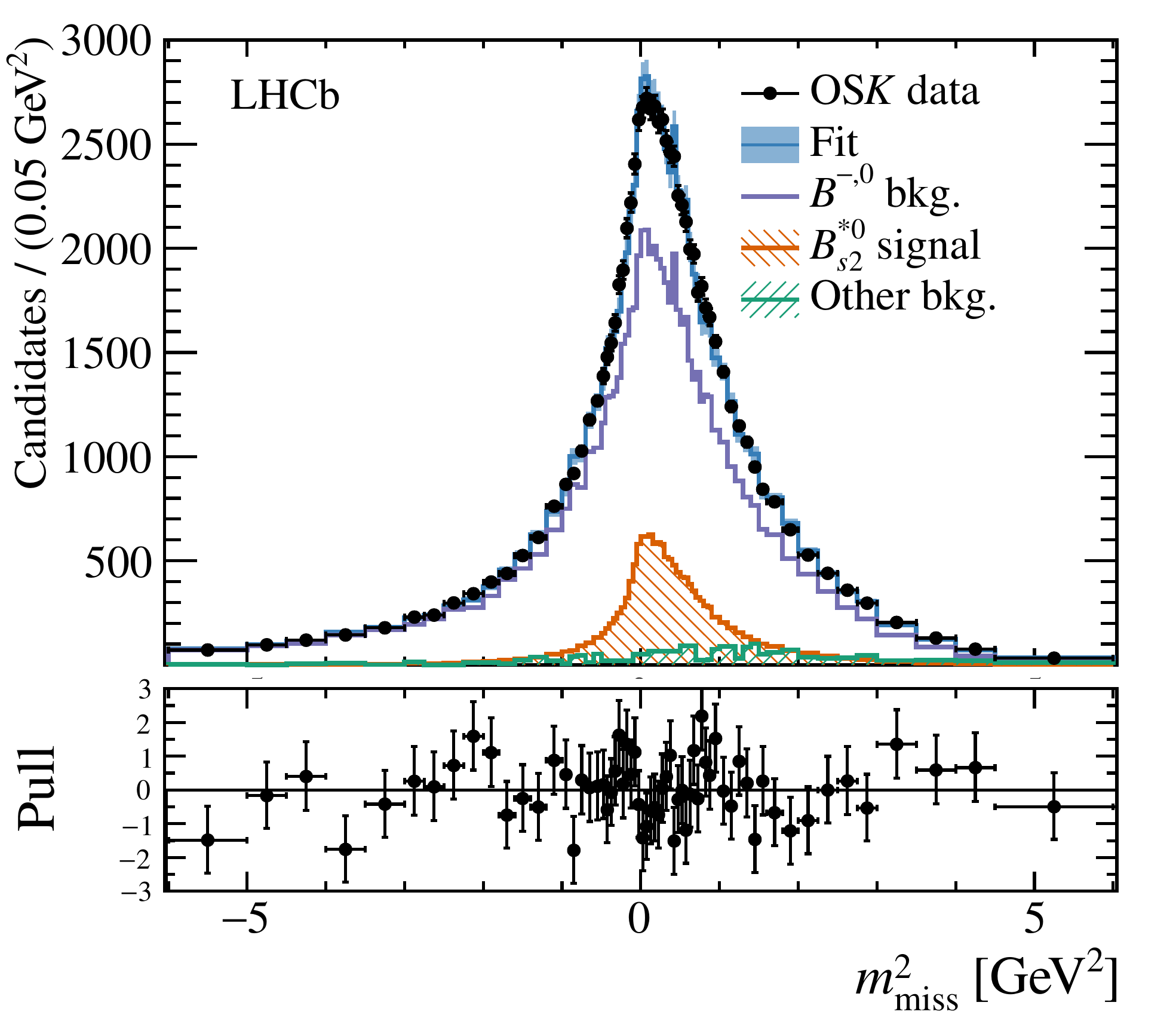}
  \end{subfigure}
  \begin{subfigure}{0.5\textwidth}
    \includegraphics[width=\textwidth]{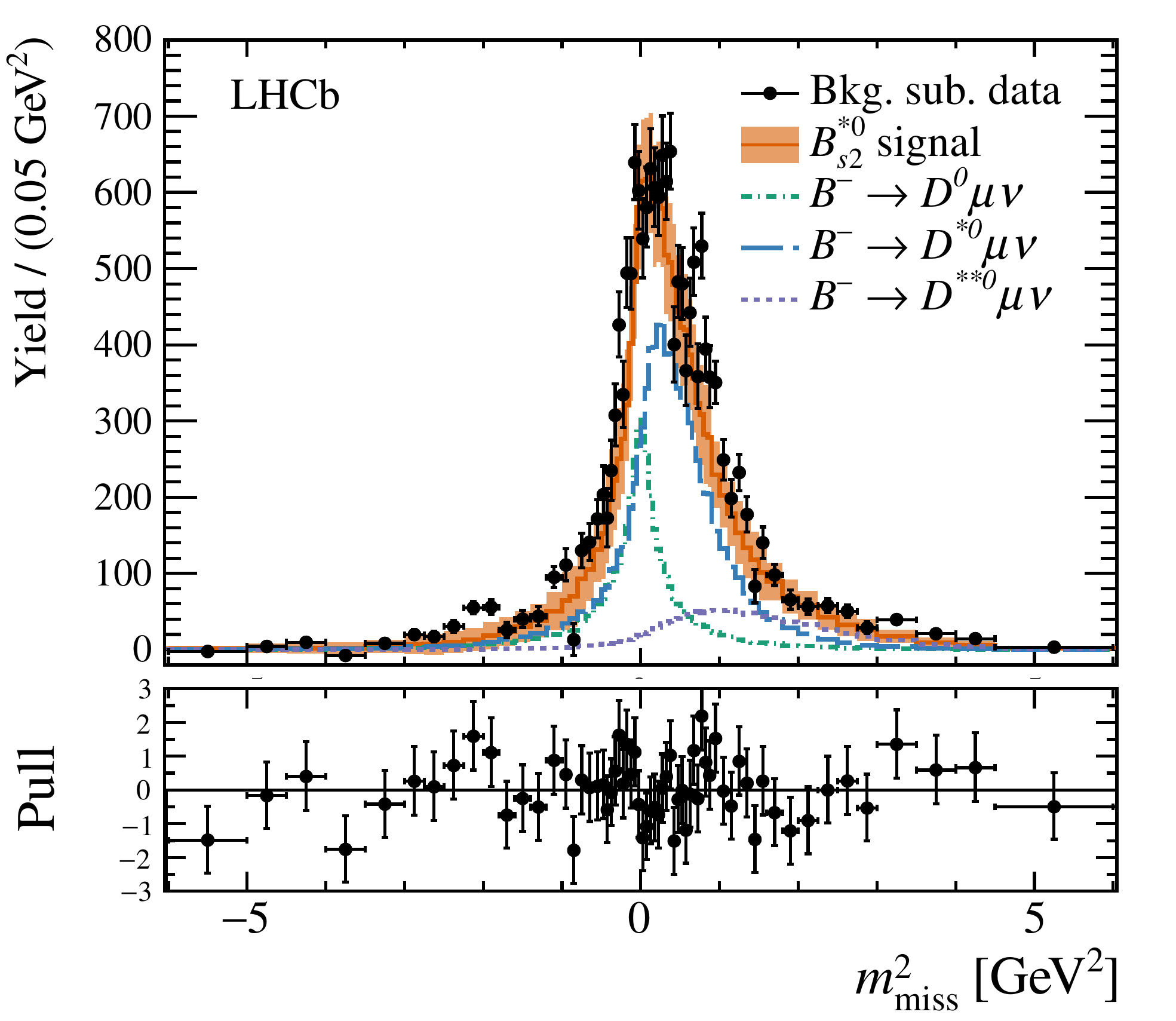}
  \end{subfigure}
  \caption{Template fit to the missing-mass distribution. The nuisance parameters used to quantify the template statistical uncertainties are set to their nominal values. The full distribution (left) is shown, comparing the background to the sum of the signal templates.  The background-subtracted distribution (right) is compared to the breakdown of the signal components. The statistical uncertainty in the background templates is represented as the shaded band around the fit. In the pull distribution, the statistical uncertainty of the background templates is added to the statistical uncertainty of the data points.\label{fig:fit}}
\end{figure}

\begin{figure}[htbp]
  \begin{center}
    \includegraphics[width=0.7\textwidth]{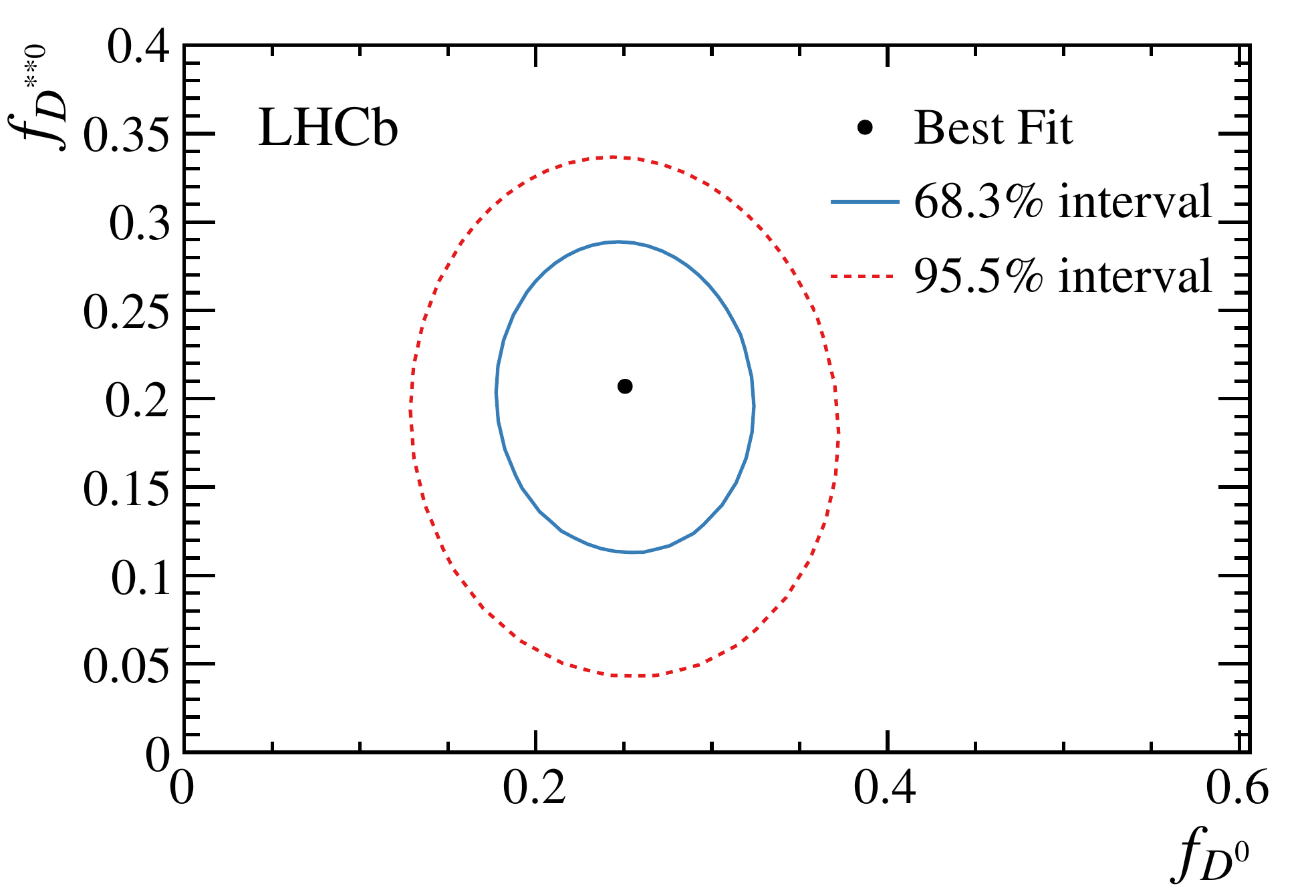}
  \end{center}
  \caption{Contours for 68.3\% and 95.5\% confidence intervals for the fractions of \decay{\Bub}{\Dz X \mun \neumb} into the exclusive \decay{\Bub}{\Dz\mun\neumb} channel and the higher excited \decay{\Bub}{\left(\decay{\dstst}{\Dz X}\right)\mun\neumb} channels. The alternate fit using different branching fractions for different \dstst states is not included. \label{fig:contour}}
\end{figure}

The result of the template fit is shown in \cref{fig:fit}.
We find the parameters of interest
\begin{align*}
  f_{\Dz} &= 0.25 \pm 0.06, \\
  f_{\dstst} &= 0.21 \pm 0.07,
\end{align*}
where the uncertainty is the total due to statistical and systematic uncertainties.  Contours for the 68.3\% and 95.5\% confidence intervals for the nominal fit are shown in \cref{fig:contour}. From the conditional covariance of the two parameters of interest combined with the fit result using alternate \dstst branching fractions, the correlation coefficient of the two parameters is $\rho = -0.38$, which is dominated by the change in the alternate branching-fraction fit. The fraction $f_{\Dstarz}$ is equal to $ 1 - f_{\Dz} - f_{\dstst} = 0.54\pm 0.07$, but this cannot be taken as an independent determination.

The results are compatible with expectations based on previous exclusive measurements~\cite{mynote}. Because of the uncertainty on the \dstst component, the results do not yet favor a particular explanation for the exclusive--inclusive gap.

We have demonstrated that the reconstruction of the momentum of \Bub decays with missing particles using \bstwostb decays is a viable method at the LHCb experiment. This technique requires much larger data sets than measurements with inclusive \Bub selections, but measuring the missing mass provides important discriminating power between different decay modes, and between signal and backgrounds. This is a promising method to employ with the additional data that the LHCb experiment has collected in Run 2 and will collect in the future.

\section*{Acknowledgements}
%
%
\noindent We express our gratitude to our colleagues in the CERN
accelerator departments for the excellent performance of the LHC. We
thank the technical and administrative staff at the LHCb
institutes.
We acknowledge support from CERN and from the national agencies:
CAPES, CNPq, FAPERJ and FINEP (Brazil); 
MOST and NSFC (China); 
CNRS/IN2P3 (France); 
BMBF, DFG and MPG (Germany); 
INFN (Italy); 
NWO (Netherlands); 
MNiSW and NCN (Poland); 
MEN/IFA (Romania); 
MinES and FASO (Russia); 
MinECo (Spain); 
SNSF and SER (Switzerland); 
NASU (Ukraine); 
STFC (United Kingdom); 
NSF (USA).
We acknowledge the computing resources that are provided by CERN, IN2P3
(France), KIT and DESY (Germany), INFN (Italy), SURF (Netherlands),
PIC (Spain), GridPP (United Kingdom), RRCKI and Yandex
LLC (Russia), CSCS (Switzerland), IFIN-HH (Romania), CBPF (Brazil),
PL-GRID (Poland) and OSC (USA).
We are indebted to the communities behind the multiple open-source
software packages on which we depend.
Individual groups or members have received support from
AvH Foundation (Germany);
EPLANET, Marie Sk\l{}odowska-Curie Actions and ERC (European Union);
ANR, Labex P2IO and OCEVU, and R\'{e}gion Auvergne-Rh\^{o}ne-Alpes (France);
Key Research Program of Frontier Sciences of CAS, CAS PIFI, and the Thousand Talents Program (China);
RFBR, RSF and Yandex LLC (Russia);
GVA, XuntaGal and GENCAT (Spain);
the Royal Society
and the Leverhulme Trust (United Kingdom);
Laboratory Directed Research and Development program of LANL (USA).

\section*{Appendix}

\appendix

\section{Derivation of the $\boldsymbol{\Bub}$ meson energy}
\label{app:eq}

Consider a known \Bub momentum direction with unknown energy and a kaon of momentum $p_K$ at an angle $\theta$ in the laboratory frame with respect to it.  Taking the \Bub direction as the $z$-axis, the squared mass of the $\Bub\Kp$ system is
\begin{align}
  m_{BK}^2 &= \qty| \mqty( E_B \\ 0 \\ 0 \\ \sqrt{E_B^2 - m_B^2}) + \mqty( \sqrt{p_K^2 + m_K^2} \\ p_K\sin\theta \\ 0 \\ p_K\cos\theta) |^2. \label{eqn:mbk} \\
\intertext{For a particular $m_{BK}$ hypothesis, \cref{eqn:mbk} can be written}
   m_{BK}^2 &= \qty(E_B + \sqrt{p_K^2 + m_K^2})^2 - p_K^2\sin^2\theta \\
   & \phantom{{}={}} - \qty( \sqrt{E_B^2 - m_B^2} + p_K\cos\theta)^2 \nonumber \\
   &= E_B^2 + 2E_BE_K +  m_K^2 + \qty(p_K^2 - p_K^2\sin^2\theta) \\
   &\phantom{{}={}}- E_B^2 + m_B^2 - 2p_K\cos\theta\sqrt{E_B^2 - m_B^2} - p_K^2\cos^2\theta. \nonumber \\
  \intertext{Rearranging terms, squaring to remove the root, and using $\Delta^2 = m_{BK}^2 - m_{B}^2 - m_K^2$ gives}
   0 &= E_B^2\qty( 4\qty(E_K^2 - p_K^2\cos^2\theta)) + E_B\qty(-4E_K\Delta^2) \\
   &\phantom{{}={}} + \qty( 4m_B^2p_K^2\cos^2\theta + \Delta^4). \nonumber
\intertext{The solution to the quadratic equation for $E_B$ is}
  E_B &= \frac{\Delta^2}{2E_K}\frac{1}{1 - \qty(p_K/E_K)^2 \cos^2\theta} \qty[ 1 \pm \sqrt{d} ],\\
  \intertext{where}
    d &= \frac{p_K^2}{E_K^2} \cos^2\theta - \frac{4m_B^2p_K^2\cos^2\theta}{\Delta^4}\qty(1 - \frac{p_K^2}{E_K^2} \cos^2\theta ).
\end{align}

\setboolean{inbibliography}{true}
\bibliographystyle{LHCb}
\bibliography{standard,d0munu,LHCb-PAPER,LHCb-CONF,LHCb-DP,LHCb-TDR}

\newpage


 
\newpage
\centerline{\large\bf LHCb collaboration}
\begin{flushleft}
\small
R.~Aaij$^{27}$,
B.~Adeva$^{41}$,
M.~Adinolfi$^{48}$,
C.A.~Aidala$^{73}$,
Z.~Ajaltouni$^{5}$,
S.~Akar$^{59}$,
P.~Albicocco$^{18}$,
J.~Albrecht$^{10}$,
F.~Alessio$^{42}$,
M.~Alexander$^{53}$
% ...

{\footnotesize \it
% ...
$ ^{5}$Clermont Universit{\'e}, Universit{\'e} Blaise Pascal, CNRS/IN2P3, LPC, Clermont-Ferrand, France\\
% ...
$ ^{10}$Fakult{\"a}t Physik, Technische Universit{\"a}t Dortmund, Dortmund, Germany\\
% ...
$ ^{18}$INFN Laboratori Nazionali di Frascati, Frascati, Italy\\
% ...
$ ^{27}$Nikhef National Institute for Subatomic Physics, Amsterdam, Netherlands\\
% ...
$ ^{41}$Instituto Galego de F{\'\i}sica de Altas Enerx{\'\i}as (IGFAE), Universidade de Santiago de Compostela, Santiago de Compostela, Spain\\
$ ^{42}$European Organization for Nuclear Research (CERN), Geneva, Switzerland\\
% ...
$ ^{48}$H.H. Wills Physics Laboratory, University of Bristol, Bristol, United Kingdom\\
% ...
$ ^{53}$School of Physics and Astronomy, University of Glasgow, Glasgow, United Kingdom\\
% ...
$ ^{59}$University of Cincinnati, Cincinnati, OH, United States\\
% ...
$ ^{61}$Syracuse University, Syracuse, NY, United States\\
% ...
$ ^{73}$University of Michigan, Ann Arbor, United States, associated to $^{61}$\\
% ...
}
\end{flushleft}



\end{document}